\documentclass[runningheads]{llncs}
\usepackage{comment}
\usepackage{cite}
\usepackage{amsmath,amssymb,amsfonts}
\usepackage{algorithmic}
\usepackage{graphicx}
\usepackage{textcomp}
\usepackage{xcolor}
\usepackage{caption}
\usepackage{subcaption}
\begin{document}
\title{
Discontinuous Galerkin schemes for 
master equations modeling open quantum systems\thanks{Supported by The University of Texas at San Antonio (UTSA).}}
\titlerunning{DG for master equations of open quantum systems}
\author{Jos\'e A. Morales Escalante\inst{1,2}\orcidID{0000-0001-6650-9510}}
\authorrunning{J. Morales Escalante}
%
\institute{Department of Mathematics, The University of Texas at San Antonio, San Antonio TX 78249, USA 
\\
\url{https://sciences.utsa.edu/faculty/profiles/morales-jose.html} \and
Department of Physics and Astronomy, The University of Texas at San Antonio, San Antonio TX 78249, USA\\
\email{jose.morales4@utsa.edu}}
\maketitle              

\begin{abstract}
This work presents a numerical analysis of a
Discontinuous Galerkin (DG) method for a
{\color{black} transformed
master equation modeling 
an open quantum system:
a
quantum sub-system interacting with
a noisy environment}. 
It is shown that the presented transformed master equation has a reduced computational cost in comparison to a Wigner-Fokker-Planck model of the same system for the general case of 
{\color{black} non-harmonic}
potentials via DG schemes. Specifics of a  
{\color{black}
Discontinuous Galerkin
(DG)
}
numerical scheme adequate for the system of convection-diffusion equations obtained for our Lindblad master equation in position basis are presented. This lets us solve computationally the transformed system of interest modeling our open quantum system problem. 
{\color{black} The benchmark case of a harmonic potential is then presented}, for which the numerical results are compared against the analytical steady-state solution of this problem. Two non-harmonic cases are then presented: the linear and quartic potentials are modeled via our DG framework, for which we show our numerical results. 

\end{abstract}

\section{Introduction}
Open quantum systems model the important physics problem of the interaction (through energy exchanges, for example) between a quantum (sub-)system and a usually larger environment. 
{\color{black}
This problem has important applications in quantum computing and information sciences, for two main reasons: first, the study of noise is fundamental in quantum computing as one of the main issues with practical implementations of these devices from an experimental point of view is their errors due to their high sensitivity to the noise introduced by the environment; second, recent studies \cite{PhysRevResearch.6.033147}
show that ground state preparation (that is, the guiding of the quantum state towards the ground state from which to start a quantum computation)
can be achieved via the introduction of Lindbladians (namely, Markovian noise).} 
{\color{black}
If a Markovian dynamics for the interaction is assumed, open quantum systems are mathematically expressed in quantum information science by 
a Lindblad
master equation
for the density matrix of the system \cite{MikeIke} to describe, for example, a noisy quantum channel. 
When the variables of interest for the problem are continuous,
such as position or momentum,
these 
}
Lindblad master equations can be converted into Wigner-Fokker-Planck (WFP) equations by applying to them a Wigner transform \cite{Ferraro, Arnold}. The density matrix is converted under this transformation into the Wigner quasi-probability density function representing the system.
The Wigner-Fokker-Planck formulation is then, mathematically, equivalent 
to the Lindblad master equation description of open quantum systems
{\color{black} when they admit a description in terms of continuous quantum variables}.
The quantum Fokker-Planck operator terms (which are the analogs in the Wigner formulation of the Lindblad
{\color{black} jump}
operators of the master equation) represent the {\color{black}
diffusion and friction in the system, 
both of these phenomena
}
due to the interaction with the environment, 
{\color{black} say}
via energy exchanges, as abovementioned.

The following dimensionless model of an open quantum system will be first considered, which is given
 by the WFP equation 
 with an arbitrary potential  as
 below \cite{Arnold,Gamba} (with choice of units such that
 $\hbar=1=m$)
\begin{equation}
w_{t}+k\cdot\nabla_{x}w+\Theta[V]{w}=Q_{FP}\{w\}, 
\end{equation}
\begin{equation}
 Q_{FP}\{w\} = 
 \nabla_{(x,k)}\cdot(D\nabla_{(x,k)} w)
 +\gamma\nabla_k\cdot(wk),
\end{equation}
where 
$Q_{FP}$
 is the quantum Fokker-Planck operator that models the interaction of the
 system with its environment (representing it in the Wigner picture by diffusive and friction terms, with diffusion matrix $D$ and friction coefficient $\gamma$). The following particular case
for the aforementioned operator will be considered, 
\[
Q_{FP}\{w\}=\nabla^2_{x}w+\nabla^2_{k}w+\nabla_{k}\cdot(wk),
\]
corresponding to an identity diffusion matrix $D=I$ and a unit friction coefficient $\gamma=1$, 
 and the non-local pseudo-differential operator related to the potential 
$V$  is given by
\[
\Theta[V]\{w\}=\frac{-i}{(2\pi)^{d}}\int_{\mathcal{\mathcal{R}}^{2d}}\delta V(x,\eta) w(x,k',t)e^{i\eta\cdot(k-k')}dk'd\eta,
\]
\[
\delta V(x,\eta)
= V(x+\eta/2)-V(x-\eta/2),
\]

 which can also be represented as 
\[
\Theta[V]\{w\}=\frac{-i}{(2\pi)^{d}}\int_{\mathcal{\mathcal{R}}^{d}}[V(x+\eta/2)-V(x-\eta/2)]\widehat{w}e^{i\eta\cdot k}d\eta,
\]
 with 
$\widehat{w}$
 the Fourier transform of the Wigner function
$w$,
\[
\widehat{w}(x,\eta,t)=\int_{\mathcal{\mathcal{R}}^{d}}w(x,k',t)e^{-i\eta\cdot k'}dk'.
\]

The pseudo-differential operator is the  computationally costliest term in a simulation of the WFP system, due to the extra integration needed to be performed in order to compute it \cite{Gamba}. However, it is clear that, if a Fourier transform is applied to it, it will amount just to a multiplication by convolution theorems. Therefore, this motivates the interest in applying a Fourier transform to the Wigner function to simply obtain a density matrix in the position basis with conveniently transformed coordinates. If a Fourier transform is then applied to the WFP system, a transformed master equation  in convenient coordinates
will be obtained,
which will simplify the expression of the 
{\color{black}
term related to the potential.
}
A similar evolution equation for a density matrix in transformed coordinates has been presented in
\cite{Juengel} in the context of electron ensembles in semiconductors. 
The above-mentioned is justified by recalling that the Wigner function is defined by the Fourier transform
below,
\begin{equation}
 w(x,k,t)=\int_{\mathcal{R}^d} \rho(x+\eta/2,x-\eta/2,t)\exp(-i\eta\cdot k)d\eta,
\end{equation}

particularly as the Fourier transform of the function

\begin{equation}
 u(x,\eta,t)=\rho(x+\eta/2,x-\eta/2,t),
\end{equation}

which is a density matrix in terms of  conveniently symmetrized position coordinates. Therefore,

\begin{equation}
 w=\widehat{u}, \quad u=\check{w}, 
\end{equation}

and then
{\color{black}
\begin{equation}
 u(x,\eta,t)=\frac{1}{(2\pi)^d} 
 \int_{\mathcal{R}^d} w(x,k,t) \exp(ik\cdot\eta)dk =
\end{equation}
\begin{equation}
=
 \frac{1}{(2\pi)^d} 
 \int_{\mathcal{R}^d} w(x,k,t)\exp(-ik\cdot[-\eta])dk
 =
 \frac{\widehat{w}(x,-\eta,t)}{(2\pi)^d}.
\end{equation}
}

So the Fourier transform above is just proportional to the density matrix evaluated at conveniently transformed position coordinates,

\begin{equation}
 {\widehat{w}(x,\eta,t)}=
 {(2\pi)^d} \rho(x-\eta/2,x+\eta/2,t).
 \end{equation}

 Different numerical methods have been used in the computational simulation of models for open quantum systems.
Regarding computational modeling of open quantum systems via Wigner functions,
stochastic methods such as Monte Carlo simulations
of Fokker-Planck equations in quantum optics have been
reported \cite{doi:10.1142/S0217984994000248}, as well as
numerical discretizations of velocities for the stationary Wigner equation\cite{RevModPhys.62.745}. However, there is an inherent stochastic error in the solution of
partial differential equations
(PDEs) by Monte Carlo methods (where this error will decrease slowly, as $N^{-1/2}$, by increasing the number of samples $N$).
{\color{black}
There is literature on the computational methods for
Wigner equations based on spectral
 methods, such as work in \cite{12c710cb-09d6-3648-9a11-c65d0b7e8f85}
 related to the Wigner numerical simulation of quantum tunneling phenomena, 
 and work in \cite{RinghoferConvergence,06e29f02-3d45-3e6b-93aa-17b8a7d51143}
 related to spectral methods for Wigner-Poisson based on spectral collocation techniques. 
 More specifically, there is work on
 computational methods for Wigner equations based on
spectral elements, as in \cite{Shao_Lu_Cai_2011}, which 
presents 
adaptive conservative cell average spectral element methods for a transient Wigner Equation in quantum transport.}
There is work as well on operator splitting methods for Wigner-Poisson \cite{55c98552-dbda-399e-bf25-fc085941dc30,doi:10.1137/0732084}, and on the semidiscrete analysis of the Wigner equation \cite{GoudonSDWigner}, as well as on 
discrete models for quantum transport in semiconductor devices in
\cite{lohrengel:hal-00210604}.
There is also work in \cite{1202616} on
quantum ensemble Monte Carlo techniques for the study of a resonant tunneling diode based on a Wigner function description. 
There is work as well on the splitting-scheme solution of a collisionless Wigner equation with a non-parabolic band profile in \cite{demeio2003splitting}.
However, the nature of the phenomena for open quantum systems claims the need for a numerical method that reflects in its nature the physics of both transport and noise 
{\color{black}
(represented by diffusion for system-environment interactions assumed to be Markovian),}
therefore the interest on Discontinuous Galerkin (DG) methods that can mimic numerically convection-diffusion problems, such as Local DG or Interior Penalty methods, for example. 
{\color{black}
There is work by \cite{liu2016entropy} 
on entropy satisfying DG methods
to model nonlinear Fokker-Planck equations. }
{\color{black} In particular, there is work}
in \cite{Gamba}, with theoretical foundations on the analysis in \cite{Riviere,IJNAM6533} (all of these works for the numerical solution of scalar functions), on an adaptable DG scheme for Wigner-Fokker-Planck as a convection-diffusion equation, where a Discontinuous Galerkin method is used for the computational modeling of the problem. The main issue of using a Wigner model for open quantum systems though is the fact that, except for the case of harmonic potentials, the computationally costliest term  in a Wigner formulation of a quantum transport problem is the pseudo-differential integral operator, whereas a simple Fourier transformation of the Wigner equation might render this term as just a simple multiplication between the related density matrix and the (possibly non-harmonic) potential, and therefore not as computationally costly anymore in a master equation setting for the density matrix. 
There are literature reports of the appliciation of Discontinuous Galerkin Methods onto Quantum-Liouville type equations \cite{GaniSchulzDGquantumLiouville}, 
and numerical simulations of the Quantum Liouville-Poisson system as well \cite{Suh1991NumericalSO}.
However, this kind of equations consider only the quantum transport part of the problem, since diffusion 
{\color{black}
does not appear}
in Liouville transport, therefore the noise due to an environment is missing to be modeled in a DG setting for Lindblad master equations up to the author's knowledge.

The 
{\color{black}
novelty and significance of this paper is that it}
provides a DG-based computational solver for the Lindblad  master equation for open quantum systems described by density matrices in position basis, whose underlying numerics reflects inherently in its methodology the convective-diffusive nature of the physical phenomena of interest, solving the resulting system of master equations by means of a Discontinuous Galerkin  method to handle both the quantum transport and the diffusive noise due to
Markovian system-environment interactions, at a less expensive computational cost than a Wigner computational model for the
{\color{black}
important physics application problem of 
open quantum systems subjected to Markovian noise in a continuous-variable description.}  

{\color{black}
 The summary of the presented work is the following:  convolution
 theorems of Fourier transforms will be used to convert the pseudo-differential operator
 into a product, transforming then the WFP into a master equation for the density matrix in symmetrized position coordinates. The transport and diffusion terms will be analyzed, and the transformed master equation will then be studied numerically via a DG method, as an
 initial boundary value problem (IBVP) in a 2D position space.
 We will show the reduction of the computational
 cost for the transformed master equation formulation 
 in comparison to WFP when both are solved through DG methodologies.
 Then, numerical convergence studies are presented for the benchmark harmonic potential problem,  for which the analytical steady-state solution is known (the harmonic oscillator case), therefore proceeding to compare it to the numerical solution obtained via the DG method. 
Finally, we will show some non-harmonic examples for which our DG solver is able to help studying the respective Lindblad master equation, namely the cases of linear and quartic potentials. 
 }
 
\section{Math Model: Transformed Master Equation for open quantum systems}

 The forward Fourier transform of the WFP equation is
 the transformed master equation below, 
{\color{black}
 \begin{equation}
\widehat{w}_{t}+i\nabla_{\eta}\cdot\nabla_{x}\widehat{w}+i\widehat{w}\cdot{\delta V}=
-\eta^{2}\widehat{w}-\eta\cdot\nabla_{\eta}\widehat{w}+\nabla^2_{x}\widehat{w}.
\label{eq:FTWFP}
 \end{equation}
}
{\color{black}
The pseudo-differential operator
over the potential $V$ has been transformed back into just a
product of functions by Fourier transforming the WFP equation.}
 The main change is that the transport term is represented now
 as a derivative of higher order ($k$ was exchanged 
 for 
$i\nabla_{\eta}$
 when transforming into the Fourier space).
If one starts with
\[
\widehat{w}_{t}+\nabla_{x}\cdot(i\nabla_{\eta}\widehat{w})+\eta\cdot\nabla_{\eta}\widehat{w}=\widehat{w}\cdot\left(\frac{\delta V}{i}-\eta^{2}\right)+\nabla_{x}^{2}\widehat{w},
\]

it can be noticed that the transport term involves a complex number, so
$\widehat{w}=R+iI$
will be decomposed 
 into real and imaginary parts.
{\color{black} To explain the motivation of this decomposition, one cannot understand or express a complex-valued transport in a real-valued space such as 
$\mathbb{R}^n$. Therefore, it is indeed necessary to decompose the complex-transport term into real and imaginary parts 
  to rather better understand  this term
  and quantify it in our real-valued space. So we proceed to decompose and we get
} 
\[
R_{t}+iI_{t}+\nabla_{x}\cdot\nabla_{\eta}[-I+iR]+\eta\cdot\nabla_{\eta}[R+iI]=
\]
\[
[R+iI]\cdot\left(\frac{\delta V}{i}\right)-[R+iI]\cdot\left(\eta^{2}\right)+\nabla_{x}^{2}[R+iI].
\]

Let's focus now on the benchmark case of dimension
$d=1$
{\color{black}
for the sake of the concrete discussion of a particular dimensional case.
}
 {\color{black}
The}
Laplacian reduces to 
$\partial_{x}^{2}$,
 so we have the following system
 {\color{black}
\[
R_{t}+(0,\eta)\cdot(\partial_{x}R,\partial_{\eta}R)=I\left({\delta V}\right)
-R\eta^{2}+\partial_{x}^{2}R+\partial_{x\eta}I,
\]

\[
I_{t}+(0,\eta)\cdot(\partial_{x}I,\partial_{\eta}I)=-R\cdot\left({\delta V}\right)
-I\eta^{2}+\partial_{x}^{2}I-\partial_{x\eta}R.
\]
}
 The system above has the transport terms on the left-hand side. The
 ``source'', decay, and diffusive terms represented by second-order partials are on the
 right-hand side.
 {\color{black}
 If the transport and diffusion are both to be expressed in divergence form, one can pass an
 extra term to the other side, rendering 
\[
R_{t}+\partial_{\eta}(\eta R)=I\left({\delta V}\right)
+(1-\eta^{2})R +
\partial_{(x,\eta)}\cdot \left(
A\left(
\begin{matrix}
\partial_{x} R \\
\partial_{\eta}R
\end{matrix}
\right)
+
B\left(
\begin{matrix}
\partial_{x} I \\
\partial_{\eta}I
\end{matrix}
\right)
\right),
\]

\[
I_{t}+\partial_{\eta}(\eta I)=-R\cdot\left({\delta V}\right)
+(1-\eta^{2})I +
\partial_{(x,\eta)}\cdot \left(
A\left(
\begin{matrix}
\partial_{x} R \\
\partial_{\eta}R
\end{matrix}
\right)
-
B\left(
\begin{matrix}
\partial_{x} I \\
\partial_{\eta}I
\end{matrix}
\right)
\right),
\]
}

{\color{black}
with
\[
 A =
 \left(\begin{array}{cc}
1 & 0\\
0 & 0
\end{array}\right),
\quad 
B =
 \left(\begin{array}{cc}
0 & 1/2\\
1/2 & 0
\end{array}\right).
\]
}

 These convective-diffusive systems (where the transport  is only on 
$\eta$) can be expressed in matrix form.
 The matrix system in gradient form is 
\[
\partial_{t}\left(\begin{array}{c}
R\\
I
\end{array}\right)+\eta\partial_{\eta}\left(\begin{array}{c}
R\\
I
\end{array}\right)=-\left(\begin{array}{cc}
\eta^{2} & -{\delta V}\\
{\delta V} & \eta^{2}
\end{array}\right)\left(\begin{array}{c}
R\\
I
\end{array}\right)
\]
\[
+
\left(\begin{array}{cc}
\partial_{x} & \partial_{\eta}\\
-\partial_{\eta} & \partial_{x}
\end{array}\right)\left(\begin{array}{c}
\partial_{x}R\\
\partial_{x}I
\end{array}\right).
\]
{\color{black} It}
 is evident the left-hand side has a convective
 transport structure and the right-hand side has matrix terms related to decay, source (partly due to the
 potential), and diffusive behavior.
 If we define 
$
 u=(R,I)^T
$
  we can express the system in gradient form as
\begin{equation} \label{eq:gradform1}
 \partial_{t}u+\partial_{(x,\eta)}u\cdot
\left(\begin{array}{c}
0\\
\eta
\end{array}\right)
=
\left(\begin{array}{cc}
\partial_{x} & \partial_{\eta}\\
-\partial_{\eta} & \partial_{x}
\end{array}\right)\partial_{x}u
-\left(\begin{array}{cc}
\eta^{2} & -{\delta V}\\
{\delta V} & \eta^{2}
\end{array}\right)u,
\end{equation}

{\color{black}
where Equation (\ref{eq:gradform1}) above indicates that the transport is only vertical, and the gradient in  the diffusion term 
$\partial_x u$}
is related to $x$-partials.

 \section{Methodology: DG Formulation for Master Equations in transformed position coordinates}

 A DG scheme for a vector variable will be presented for our master equation (where naturally the test function
 will be a vector too), and where we will perform an inner product multiplication between these vector functions
 to obtain the weak form for our related DG methodology.

\subsection{DG method for a transformed Master Equation}

{\color{black}
One can observe from the development of the section above that our problem has the form of a convection-diffusion equation even if for a vector-valued function. As was mentioned in the Introduction, DG methods are designed specifically to reflect in the numerics the physics of the problem, either being hyperbolic or of convective-diffusive nature.
The original DG methodology for hyperbolic problems is modified 
to consider diffusion with a DG framework as well for elliptic equations \cite{Riviere} and convection-diffusion ones \cite{IJNAM6533}. 
A penalty term is then included in approaches such as \cite{Riviere,IJNAM6533,Gamba} to control the issue of second derivatives of discontinous basis functions. In this way, the DG method captures the physics of our quantum transport and includes the diffusion representing noise (taking care of avoiding singularities introduced by it through a penalty term).  
}

Below one proceeds to describe then how to solve the system resulting from a transformed master equation, in convenient position coordinates, by means of a DG method, as in \cite{IJNAM6533, Gamba} for Wigner-Fokker-Planck in particular and convection-diffusion equations (scalar equations for both references mentioned), respectively. 
In this type of DG method, some penalty terms are introduced in order to account for the diffusion when treated by DG methodologies. More information can be found in \cite{IJNAM6533, Gamba}.

The DG formulation (at the semi-discrete level) for the vector-valued master equation system is then the following:
 Find 
$u=\left(R,I\right)^T$
{\color{black}
$\in V_h$ in a vector trial space}
 such that, for all test functions 
$v=\left(w,z\right)^T$
{\color{black}
$\in V_h$ in the vector test space
(where the test and trial spaces are the same, such that the function components belong to the piecewise polynomial space $P_\kappa$ of degree $\kappa$)
}
the following holds,
\begin{equation}
\partial_t (v,u) = a(v,u) ,
\end{equation}
with the bilinear form $a(u,v)$ (including also penalty terms) given by

\begin{eqnarray}
a(u,v) &=& (\nabla w, A\nabla R - bR) + 
(\nabla z, A\nabla I - b I)
+(\nabla w, B\nabla I) - (\nabla z, B\nabla R) \nonumber\\
&&-(w,I\delta V) + (w,R \eta^2)
+(z,R\delta V) + (z,I \eta^2) \nonumber\\
&& + (\alpha/h)\langle[w],A[R ]\rangle
+ (\alpha/h)\langle[w],B[I]\rangle \\
&&+ (\alpha/h)\langle[z],A[I]\rangle
- (\alpha/h)\langle[z],B[R]\rangle \nonumber\\
&&-\langle[w],A \{\nabla R\}\cdot n\rangle
-\langle[w],B \{\nabla I\}\cdot n \rangle 
-\langle[z],A \{\nabla I\}\cdot n\rangle
\nonumber\\
&&-\langle [R], A\{\nabla w\}\cdot n\rangle
-\langle [I], B\{\nabla w\}\cdot n\rangle \nonumber\\
&&+\langle [R], B\{\nabla z\} \cdot n\rangle
-\langle [I], A\{\nabla z\}\cdot n\rangle
+\langle[z],B \{\nabla R\}\cdot n \rangle \nonumber\\
&&+\langle[w],[\widehat{b} R]\rangle + \langle w, \widehat{b} R\rangle 
+\langle[z],[\widehat{b} I]\rangle + \langle z, \widehat{b} I\rangle, \nonumber
\end{eqnarray}

where $[\cdot]$ stands for jumps,  $\{\cdot\}$ stands for averages, $(\cdot,\cdot)$ stands for volume integrals, $\langle\cdot,\cdot\rangle$ stands for surface integrals, $n$ stands for outward unit normals, $b=(0,\eta)^T$ stands for the transport vector in our 2D position domain, and 
$\widehat{b}=(b\cdot n + |b\cdot n|)/2$ is related to the upwind flux rule.
{\color{black}
We define as well the matrices representing the diffusive nature of our problem in its real-valued representation as before, namely,
$$A=\begin{pmatrix}
1 & 0 \\
0 & 0
\end{pmatrix}, \quad
B=\begin{pmatrix}
0 & 1/2 \\
1/2 & 0
\end{pmatrix},$$
as well as the objects
$\nabla=(\partial_x,\partial_\eta)$,
$\alpha$ being a penalty parameter (which we take to be $\alpha=1$), and
$h$ being the diameter of the mesh
(the largest possible distance between two points inside any element in the domain). 
}

We discretize
the time evolution by applying an implicit theta method. In order to advance
from $u_0$ to $u$ in the next time-step, the method below is then used,

\begin{equation}
(v,u-u_0)/\Delta t = \theta a(v,u) +(1-\theta)a(v,u_0),   
\end{equation}

where a value of 
$\theta=1/2\in[0,1]$
is chosen, since it is known that this 
value, among all possible in [0,1],
gives the highest order of convergence for implicit time evolution methods
\cite{Iserles}.

\section{
Computational Cost of Master Equations vs WFP via DG methods
}

Our particular case of the WFP equation
{\color{black}
(normalized
as in \cite{Gamba} by taking an identity diffusion matrix
$D=I_d$, a friction coefficient of $\gamma=1$, and a spring constant for the harmonic potential such that
the frequency is $\omega_0=1$) 
}
has the form 
\[
w_{t}+k\cdot\nabla_{x}w-
\]

\[
\frac{i}{(2\pi)^{d}}\int_{\mathcal{\mathcal{R}}^{2d}}[V(x+\frac{\eta}{2})-V(x-\frac{\eta}{2})]w(x,k',t)e^{i\eta\cdot(k-k')}dk'd\eta
\]

\[
=\nabla_{k}\cdot(kw)+\Delta_{x}w+\Delta_{k}w.
\]

 If the size of the global basis for 
$w(x,k,t)=\sum_{i=1}^{n}\sum_{j=1}^{m}\sum_{p=0}^{P}\chi_{ij}(x,k)c_{p}^{ij}(t)\phi_{p}^{ij}(x,k)$
 is 
$N=(P+1)nm$, where 
$P+1$
 is the dimension of our local polynomial basis
 {\color{black} for the trial space
 $\left\{\phi_p^{ij}(x,k)\right\}_{p=0}^{P}$,}
$n$
 the number of intervals in 
$x$, 
$m$
 the number of intervals in 
$k$, the dimension of the test space 
$\mathrm{span}\{v_{k}(x,k)\}_{k=1}^{N}=\mathrm{span}\{\{\{v_{p}^{ij}(x,k)\}_{p=0}^{P}\}_{i=1}^{n}\}_{j=1}^{m}$
 will be the same, and an DG formulation for the Wigner-Fokker-Planck
 equation would look like (
$1\leq q\leq N$
)

\[
\sum_{i=1}^{n}\sum_{j=1}^{m}\sum_{p=0}^{P}(v_{q}^{ij}|\phi_{p}^{ij})\frac{dc_{p}^{ij}}{dt}\chi_{ij}-\sum_{i=1}^{n}\sum_{j=1}^{m}\sum_{p=0}^{P}(\nabla_{x}v_{q}^{ij}|k\phi_{p}^{ij})c_{p}^{ij}\chi_{ij}-\sum_{i=1}^{n}\sum_{j=1}^{m}\sum_{p=0}^{P}(\nabla_{k}v_{q}^{ij}\cdot k|\phi_{p}^{ij})c_{p}^{ij}\chi_{ij}
\]

\[
+\sum_{i=1}^{n}\sum_{j=1}^{m}\sum_{p=0}^{P}\left\langle [v_{q}^{ij}]|[k\phi_{p}^{ij}]\right\rangle c_{p}^{ij}\chi_{ij}+\sum_{i=1}^{n}\sum_{j=1}^{m}\sum_{p=0}^{P}\left\langle v_{q}^{ij}|k\phi_{p}^{ij}\right\rangle c_{p}^{ij}\chi_{ij}
\]

\[
+(v_{q}^{ij}|\int_{\mathcal{\mathcal{R}}^{2d}}\delta V(x,\eta)\phi(x,k')\frac{e^{i\eta\cdot(k-k')}}{i(2\pi)^{d}}dk'd\eta)c_{p}^{ij}+
\]

\[
=-\sum_{i=1}^{n}\sum_{j=1}^{m}\sum_{p=0}^{P}(\nabla_{x}v_{q}^{ij}|\nabla_{x}\phi_{p}^{ij}c_{p}^{ij})\chi_{ij}-\sum_{i=1}^{n}\sum_{j=1}^{m}\sum_{p=0}^{P}(\nabla_{k}v_{q}^{ij}|\nabla_{k}\phi_{p}^{ij})c_{p}^{ij}\chi_{ij}+
\]

\[
+\sum_{i=1}^{n}\sum_{j=1}^{m}\sum_{p=0}^{P}(\alpha/h)\langle[v_{q}^{ij}n]|[c_{p}^{ij}\phi_{p}^{ij}n]\rangle-\sum_{i=1}^{n}\sum_{j=1}^{m}\sum_{p=0}^{P}\langle[v_{q}^{ij}n]|\{\nabla\phi_{p}^{ij}c_{p}^{ij}\}\rangle-\sum_{i=1}^{n}\sum_{j=1}^{m}\sum_{p=0}^{P}\langle\{\nabla v_{q}^{ij}\}|[\phi_{p}^{ij}a_{p}^{ij}n]\rangle
\]

so we have, at the semi-discrete level, 
$11N$
 operations related to matrix-vector multiplications, but we must consider
 the cost of computing the matrix elements.
 If we go into the details of the cost of each term, assuming we use a piece-wise polynomial basis as traditionally used in DG (nonzero only inside a given
 element) we will see that computing all terms but the pseudo-differential
 operator takes work of 
$O(10nm[P+1]^{2})$
 integrations (due to the piece-wise nature of the spaces).
 Specifically, without considering the pseudo-differential term, we need
$5nm[P+1]^{2}$
 volume integrations and 
$\text{\ensuremath{5nm[P+1]^{2}}}$
 surface integrals The pseudo-differential operator involves an integral
 over the phase space where the potential and the real and imaginary parts
 of the complex exponential are involved, 
\[
\sum_{r=1}^{n}\sum_{s=1}^{m}\sum_{p=0}^{P}(v_{q}^{ij}(x,k)|\int_{\eta_{r-}}^{\eta_{r+}}\int_{k_{s-}}^{k_{s+}}\delta V\phi_{p}^{rs}(x,k')\frac{e^{i\eta\cdot(k-k')}}{i(2\pi)^{d}}dk'd\eta)c_{p}^{rs}\chi_{rs},
\]
due to the non-local nature of the double integrals, so there are possible
 extra overlaps in comparison to the other terms.
 Due to the nature of the basis, namely, piece-wise functions constituted
 by polynomials multiplied by characteristic functions, the overlap in 
$x$
 must happen in order to get nonzero terms, so only 
$i=r$
 will contribute, but the overlap in 
$k$
 will happen as a must in any case since the integration is over all momentum
 regions.
 Our integral reduces to 
\[
\sum_{r=1}^{n}\sum_{s=1}^{m}\sum_{p=0}^{P}(v_{q}^{ij}\chi^{ij}|\int_{\eta_{r-}}^{\eta_{r+}}\int_{k_{s-}}^{k_{s+}}\delta V\phi_{p}^{rs}(x,k')\frac{e^{i\eta\cdot(k-k')}}{i(2\pi)^{d}}dk'd\eta)c_{p}^{rs}\chi_{rs}
\]

\[
=\sum_{s=1}^{m}\sum_{p=0}^{P}(v_{q}^{ij}\chi^{j}(k)|\int_{\eta_{i-}}^{\eta_{i+}}\int_{k_{s-}}^{k_{s+}}\delta V\phi_{p}^{is}(x,k')\frac{e^{i\eta\cdot(k-k')}}{i(2\pi)^{d}}dk'd\eta)c_{p}^{rs}\chi_{rs},
\]

so the computational cost of this term is in principle 
{\color{black}
\[
nm^{2}[P+1]^{2}=
([P+1]nm)^{2}/n=
N^{2}/n=
(nm[P+1]^{2})m,
\]
}
 which is 
$m$
 times the usual cost of the other terms.
 This proves that the pseudo-differential term is the computationally costliest
 in the WFP numerics.
 
On the other hand, the computational cost of our transformed master equation
 in convenient coordinates can be analyzed by considering
\[
u=(R,I)=\sum_{i=1}^{n}\sum_{j=1}^{m}\sum_{p=0}^{P}\chi_{ij}(a_{p}^{ij}(t)\phi_{p}^{ij},d_{p}^{ij}(t)\phi_{p}^{ij})
\]
 and 
\[
v_{q}^{ij}=(w_{q}^{ij},z_{q}^{ij})
\]
 with equivalent trial/test spaces (respectively)
\[
\mathrm{span}\{\phi_{p}^{ij}\}_{i=1,j=1,p=0}^{i\leq n,j\leq m,p\leq P},
\]

\[
\mathrm{span}\{v_{p}^{ij}\}_{i=1,j=1,p=0}^{i\leq n,j\leq m,p\leq P}
\]

for this system formulation.
 We have then that the DG weak form can be expressed as 
\[
\sum_{i=1}^{n}\sum_{j=1}^{m}\sum_{p=0}^{P}(v_{q}^{ij}|\phi_{p}^{ij})\frac{da_{p}^{ij}}{dt}\chi_{ij}+\sum_{i=1}^{n}\sum_{j=1}^{m}\sum_{p=0}^{P}(v_{q}^{ij}|\phi_{p}^{ij})(d_{p}^{ij})'\chi_{ij}=\sum_{i=1}^{n}\sum_{j=1}^{m}\sum_{p=0}^{P}(\nabla w_{q}^{ij}|A\nabla\phi_{p}^{ij}a_{p}^{ij}-ba_{p}^{ij}\phi_{p}^{ij})
\]

\[
+\sum_{i=1}^{n}\sum_{j=1}^{m}\sum_{p=0}^{P}(\nabla z_{q}^{ij}|A\nabla\phi_{p}^{ij}d_{p}^{ij}-bd_{p}^{ij}\phi_{p}^{ij})+\sum_{i=1}^{n}\sum_{j=1}^{m}\sum_{p=0}^{P}(\nabla w_{q}^{ij}|B\nabla\phi_{p}^{ij}d_{p}^{ij})-\sum_{i=1}^{n}\sum_{j=1}^{m}\sum_{p=0}^{P}(z_{q}^{ij}|B\nabla\phi_{p}^{ij}a_{p}^{ij})
\]

\[
+\sum_{i=1}^{n}\sum_{j=1}^{m}\sum_{p=0}^{P}(w_{q}^{ij}|d_{p}^{ij}\phi_{p}^{ij}\delta V)+\sum_{i=1}^{n}\sum_{j=1}^{m}\sum_{p=0}^{P}(w_{q}^{ij}|a_{p}^{ij}\phi_{p}^{ij}\eta^{2})+\sum_{i=1}^{n}\sum_{j=1}^{m}\sum_{p=0}^{P}(z_{q}^{ij}|a_{p}^{ij}\phi_{p}^{ij}\delta V)+(z_{q}^{ij}|d_{p}^{ij}\phi_{p}^{ij}\eta^{2})
\]

\[
+\sum_{i=1}^{n}\sum_{j=1}^{m}\sum_{p=0}^{P}(\alpha/h)\langle[w_{q}^{ij}n]|A[a_{p}^{ij}\phi_{p}^{ij}n]\rangle+\sum_{i=1}^{n}\sum_{j=1}^{m}\sum_{p=0}^{P}(\alpha/h)\langle[w_{q}^{ij}n]|B[d_{p}^{ij}\phi_{p}^{ij}n]\rangle
\]

\[
+\sum_{i=1}^{n}\sum_{j=1}^{m}\sum_{p=0}^{P}(\alpha/h)\langle[w_{q}^{ij}n]|A[a_{p}^{ij}\phi_{p}^{ij}n]\rangle-\sum_{i=1}^{n}\sum_{j=1}^{m}\sum_{p=0}^{P}\langle[w_{q}^{ij}n]|A\{\nabla\phi_{p}^{ij}a_{p}^{ij}\}\rangle-\sum_{i=1}^{n}\sum_{j=1}^{m}\sum_{p=0}^{P}\langle[w_{q}^{ij}n]|B\{\nabla\phi_{p}^{ij}a_{p}^{ij}\}\rangle
\]

\[
-\sum_{i=1}^{n}\sum_{j=1}^{m}\sum_{p=0}^{P}\langle\{\nabla w_{q}^{ij}\}|A[a_{p}^{ij}\phi_{p}^{ij}n]\rangle-\sum_{i=1}^{n}\sum_{j=1}^{m}\sum_{p=0}^{P}\langle\{\nabla w_{q}^{ij}\}|B[d_{p}^{ij}\phi_{p}^{ij}n]\rangle+\sum_{i=1}^{n}\sum_{j=1}^{m}\sum_{p=0}^{P}(\alpha/h)\langle[z_{q}^{ij}n]|A[d_{p}^{ij}\phi_{p}^{ij}n]\rangle
\]

\[
-\sum_{i=1}^{n}\sum_{j=1}^{m}\sum_{p=0}^{P}(\alpha/h)\langle[z_{q}^{ij}n]|B[a_{p}^{ij}\phi_{p}^{ij}n]\rangle-\sum_{i=1}^{n}\sum_{j=1}^{m}\sum_{p=0}^{P}\langle[z_{q}^{ij}n]|A\{\nabla\phi_{p}^{ij}d_{p}^{ij}\}\rangle+\sum_{i=1}^{n}\sum_{j=1}^{m}\sum_{p=0}^{P}\langle[z_{q}^{ij}n]|B\{\nabla\phi_{p}^{ij}a_{p}^{ij}\}\rangle
\]

\[
-\sum_{i=1}^{n}\sum_{j=1}^{m}\sum_{p=0}^{P}\langle\{\nabla z_{q}^{ij}\}|A[\phi_{p}^{ij}d_{p}^{ij}n]\rangle+\sum_{i=1}^{n}\sum_{j=1}^{m}\sum_{p=0}^{P}\langle\{\nabla z_{q}^{ij}\}|B[a_{p}^{ij}\phi_{p}^{ij}n]\rangle+\sum_{i=1}^{n}\sum_{j=1}^{m}\sum_{p=0}^{P}\langle[w_{q}^{ij}]|[\widehat{b}a_{p}^{ij}\phi_{p}^{ij}]\rangle
\]

\[
+\sum_{i=1}^{n}\sum_{j=1}^{m}\sum_{p=0}^{P}\langle w_{q}^{ij}|\widehat{b}a_{p}^{ij}\phi_{p}^{ij}\rangle+\sum_{i=1}^{n}\sum_{j=1}^{m}\sum_{p=0}^{P}\langle[z_{q}^{ij}]|[\widehat{b}d_{p}^{ij}\phi_{p}^{ij}]\rangle+\sum_{i=1}^{n}\sum_{j=1}^{m}\sum_{p=0}^{P}\langle z_{q}^{ij}|\widehat{b}d_{p}^{ij}\phi_{p}^{ij}\rangle.
\]

So we have 
$29*N$
 matrix-vector multiplications-related operations, but to compute the matrices
 we simply need 
$12nm[P+1]^{2}$
 volume integrations and 
$\text{\ensuremath{17nm[P+1]^{2}}}$
 surface integrals, because we don't have any terms that involve any convolution
 or other extra integrations.
 The difference in the cost versus the WFP computation regarding integrations
 depends on the sign of 
$(12nm-5nm)+(17nm-5nm)-nm^{2}=(7+12-m)nm$
.
 Therefore, unless there's a coarse meshing in 
$k$
 for which 
$m\leq19$
, 
$m>19$
 would hold, and regarding the matrix elements computations (which involve
 the most number of operations) then the cost of our master equation will
 be more efficient than the one for WFP.

\section{Numerical Results} 

We show in this paper results for two different problems: the benchmark problem of a harmonic potential and then the case of a linear potential. 
First, a
benchmark for our WFP numerical solver 
(developed by using FEniCS \cite{AlnaesEtal2015}, 
\cite{AlnaesEtal2014}, \cite{Kirby2004}, \cite{kirby2010},
\cite{KirbyLogg2006},
\cite{LoggEtal2012}, \cite{LoggWells2010}, \cite{LoggEtal_10_2012},  \cite{LoggEtal_11_2012}, 
 \cite{BasixJoss},  \cite{ScroggsEtal2022}, \cite{OlgaardWells2010},
 {\color{black}{\cite{HronKarlin}}}
 )
is presented,  against a problem for which the analytical form of the steady state solution is known: one where the potential is harmonic, specifically $V(x)=x^2/2$. For this case it is known that the steady state solution to the WFP problem is \cite{605aac607d794ea39022c2cd16fd7cc8, Gamba}

\begin{equation}
 \mu(x,k)= \frac{\exp\left(-(\frac{|x|^2}{5}+\frac{x\cdot k}{5}+\frac{3|k|^2}{10})\right)}{2\pi\sqrt{5}}, \, (x,k)\in \mathbb{R}^{2d}. 
\end{equation}

The respective density matrix (in convenient position coordinates) for this steady state solution has the following real and imaginary components, 

\begin{eqnarray}
 u_{\mu}(x,\eta) = \rho_{\mu}(x+\eta/2,x-\eta/2) = R_{\mu}+iI_{\mu}, \\
 R_{\mu}(x,\eta,t)=\frac{e^{-(x/\sqrt{6})^{2}-(\sqrt{5}\eta/2)^{2}}}{\sqrt{6\pi}}\cos(\frac{x}{\sqrt{6}}\cdot\eta) \\ 
 I_{\mu}(x,\eta,t)=\frac{e^{-(x/\sqrt{6})^{2}-(\sqrt{5}\eta/2)^{2}}}{\sqrt{6\pi}}\sin(\frac{x}{\sqrt{6}}\cdot\eta).
\end{eqnarray}

The initial condition for our benchmark problem will be taken as the groundstate of the harmonic oscillator, whose Wigner function is 

\begin{equation}
 W_{0}(x,p)=\frac{2}{h}\exp(-a^{2}p^{2}/\hbar^{2}-x^{2}/a^{2}),
\end{equation}

and whose density matrix representation in convenient position coordinates is 

\begin{equation}
 \widehat{w}_{0}(x,\eta,t)=\frac{2\sqrt{\pi}}{ha}\exp(-\frac{x^{2}+(\eta/2)^{2}}{a^{2}}),
\end{equation}

which only has a real component (its imaginary part is zero). 
Under the environment noise, it will be deformed into the aforementioned steady state solution.

Plots of the real and imaginary components of our density matrix at the initial time
are  presented below, 
corresponding to the groundstate of the harmonic oscillator in convenient position coordinates.

\begin{figure}[h]
\centering
\includegraphics[width=0.5\textwidth]{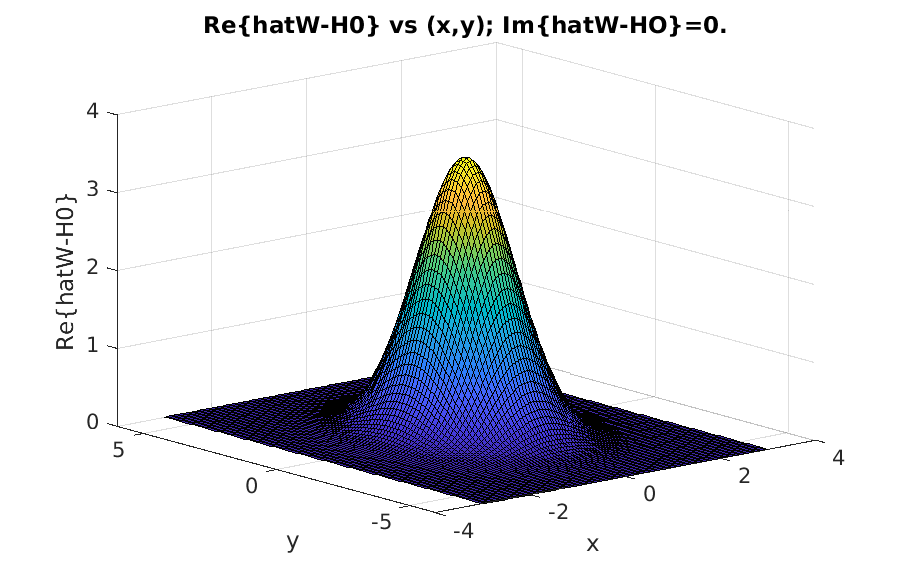}
\caption{Plot of the real component of the density matrix (in convenient position coordinates) of the harmonic oscillator groundstate (the imaginary component is zero and therefore omitted).}
\end{figure}

\begin{figure}[h]
\centering
\includegraphics[width=0.5\textwidth]{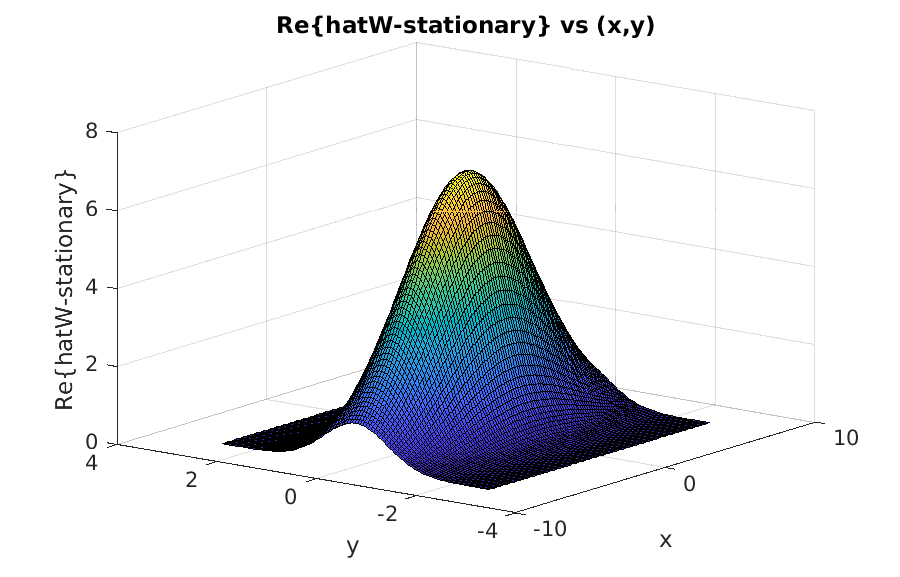}
\caption{Plot of the real component of the density matrix (in convenient position coordinates) of the steady state for our transformed master equation under a harmonic potential.}
\end{figure}

The results of our numerical simulations for the time evolution of our transformed master equations by the NIPG DG method are shown below. The time evolution is handled via a theta-method with $\theta=1/2$, as previously indicated.
{\color{black}
Dirichlet boundary conditions (BC) of two kinds were used for the benchmarking. First, BC  imposed were  related to the known analytical
{\color{black} steady-state}
solution for the case of a harmonic potential in the numerical solution of this convection-diffusion system. If the domain size is increased, these boundary conditions will converge to homogeneous BC due to Gaussian decay. This is why the second kind of Dirichlet BC used were of the homogeneous type. We studied then the difference in the performance between the two Dirichlet BC for the benchmark problem of a harmonic potential, namely studying the convergence to the analytically known steady state solution in this case through an $L_2$ error. We explain more about this study below. 
}

{\color{black}
\subsection{Boundary Conditions (BC) for the Numerical Problem}

We choose boundary conditions for our problem that translate adequately to the Physics of our problem. We take a domain $\Omega$ big enough such that the quantum information never reaches the boundary through transport. In that case, it is clear that a Wigner formulation would render homogeneous Dirichlet boundary conditions (BC) $w(x,k,t)|_{\partial\Omega}=0$. 
In order to handle it in its Fourier analog, namely the density matrix in convenient coordinates, one should choose a domain in which the homogeneous conditions translate into a Fourier series, which then will be truncated and this cut-off/truncation condition will give rise to an analog homogeneous boundary condition in $\eta$.

 Since we will compare our results to \cite{Gamba},
 we will consider their initial condition
\[
w|_{t=0}=w_{I}(x,k)
\]
 with 
 $(x,k)\in[0,L]\times[-K,K],$
 and their BC as follows, 
\[
w(x,K,t)=0=w(x,-K,t)
\]
\[
w(0,k,t)=0=w(L,k,t).
\]
 When applying our Fourier transform, the first BC will impose a constraint
 on the admissible frequencies.
 The second BC is preserved in the Fourier transform of our Wigner function.
So we will have in our density matrix problem the initial condition 
\[
\widehat{w}|_{t=0}=\widehat{w}_{I}(x,k)
\]

 with 
 $(x,\eta)\in[0,L]\times[(-J-1/4)\frac{\pi}{K},(J+1/4)\frac{\pi}{K}],$
 by imposing a cut-off frequency 
 $\frac{J\pi}{K}$,
 and with the BC 
\[
\widehat{w}(0,k,t)=0=\widehat{w}(L,k,t).
\]

 It is only missing to be specified how the BC 
 $w(x,K,t)=0=w(x,-K,t)$
 translates into our DG numerical implementation.
 Since it implies that only a discrete set of frequencies (later on truncated) are admitted, the first implication is that one has to choose a mesh
 reflecting the admitted frequencies.

That is, our numerical domain is given by the union of cells
 
\[
[(-2J-1/2)\frac{\pi}{2K},(2J+1/2)\frac{\pi}{2K}]=\cup_{j=-2J}^{2J}[(j-1/2)\frac{\pi}{2K},(j+1/2)\frac{\pi}{2K}]
\]
 with each one of them being 
\[
[\eta_{j-1/2},\eta_{j+1/2}]=[(j-1/2)\frac{\pi}{2K},(j+1/2)\frac{\pi}{2K}]=[\eta_{j-},\eta_{j+}],\quad j=-2J,\cdots,2J,
\]
and the midpoints of all these cells being the discrete set of admitted
 frequencies (up to truncation by the cutoff frequency)
\[
\{\eta_{j}=\frac{j\pi}{2K}\}_{j=-2J}^{2J},
\]
 noticing our mesh is homogeneous, of interval length
\[
\Delta\eta_{j}=\eta_{j+1/2}-\eta_{j-1/2}=\frac{\pi}{2K}=\Delta\eta,
\]
which is the distance between interval midpoints too.

 Because no frequencies higher than 
 $\eta_{\pm J}=\pm\frac{J\pi}{2K}$
 are admitted by cutoff, then the Fourier transform at the ghost boundaries
 $\pm(J+1/4)\frac{\pi}{K}$
 should vanish. That is, our second BC by restriction of the bandwidth is
\[
\widehat{w}(x,(J+1/4)\frac{\pi}{K},t)=0=\widehat{w}(x,-(J+1/4)\frac{\pi}{K},t). 
\]

\subsection{Benchmark Problem: Harmonic Potential with Steady State BC}
}

The figures below present the projection of our initial condition (the transformed density matrix of the harmonic ground state) into our DG Finite Element (FE) space $V_h^1$ of piece-wise continuous linear polynomials, for the real and imaginary components of the transformed density matrix in a position basis.

\begin{figure}[ht]
\centering
\includegraphics[width=0.49\textwidth]{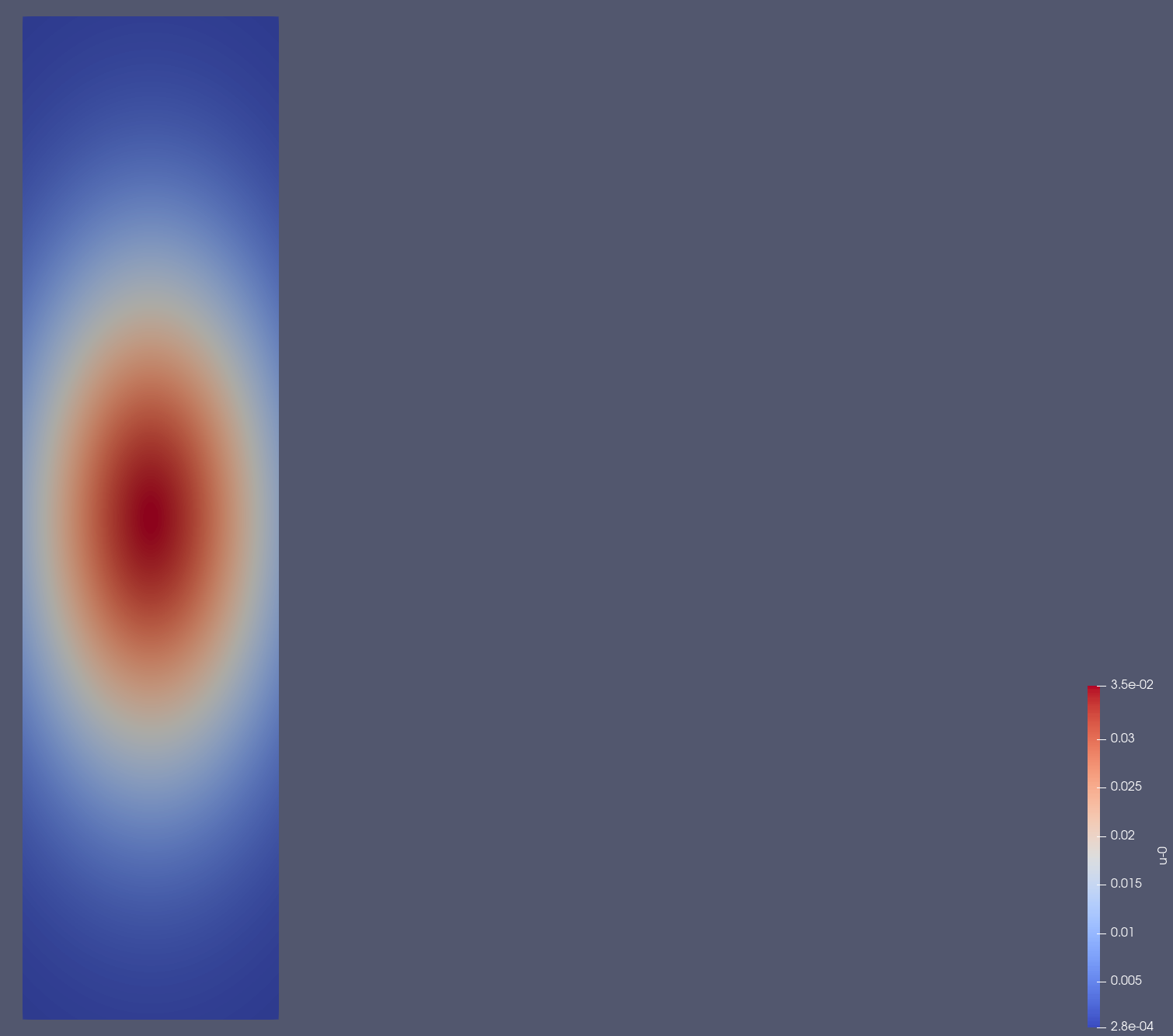}
\includegraphics[width=0.49\textwidth]{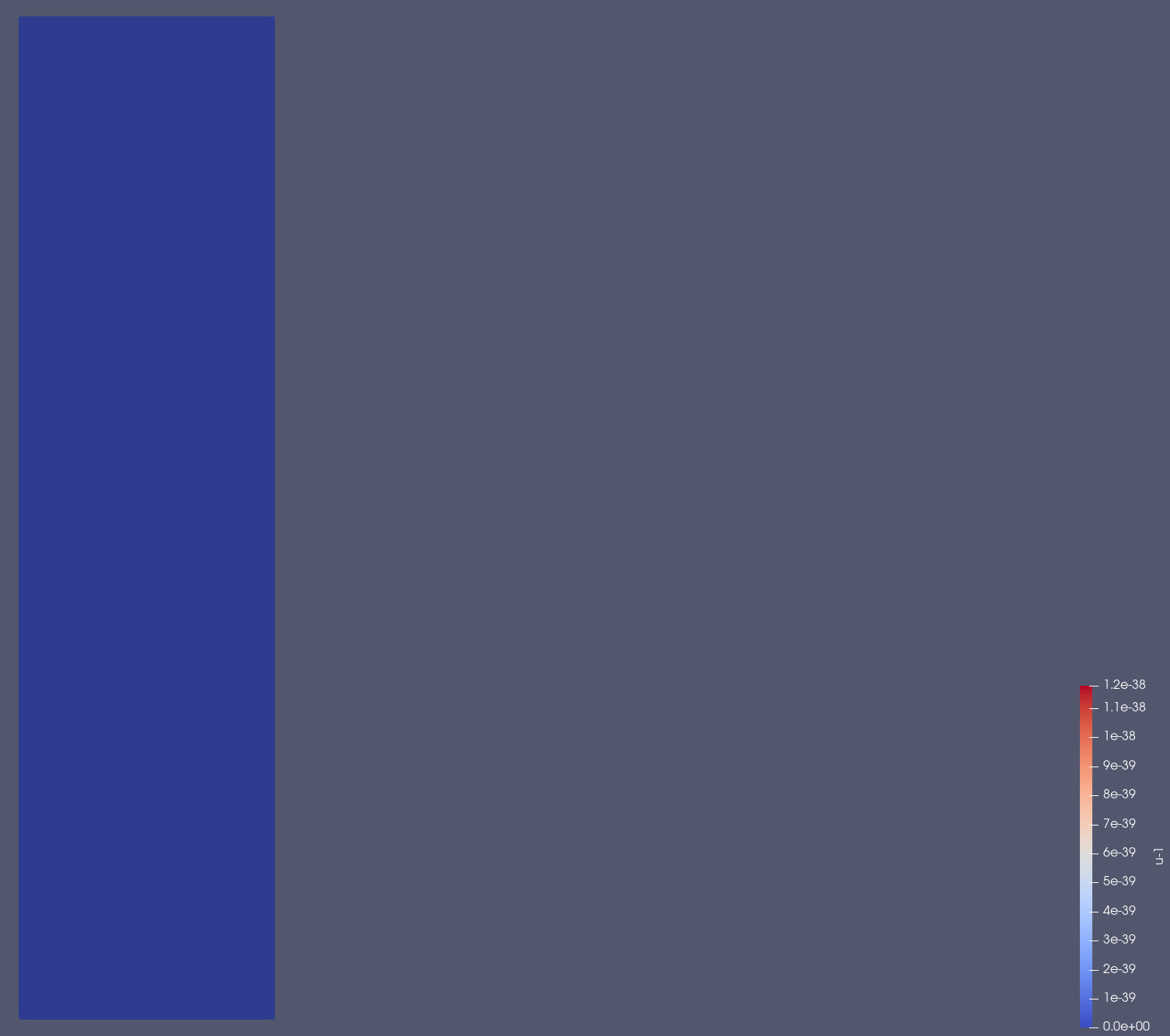}
\caption{Initial condition for the real (left) and imaginary (right) parts of the density matrix in convenient coordinates, corresponding  to a harmonic ground state (projected into the $V_h^1$
DG FE space). Remark: Color scale differs between right picture and left one. }
\end{figure}

The numerical steady-state solution for the real and imaginary components of the density matrix is presented as well, which is achieved after a long time 
(say, a physical time of $50$ in the units of the computational simulation)
under the influence of a harmonic potential. 

We first present the results for the numerical solution at $t=2$ and then at $t=50$, where in the latter case it is close to the steady state, given our convergence analysis studies to be presented below. 

\begin{figure}[ht]
\centering
\includegraphics[width=0.49\textwidth]{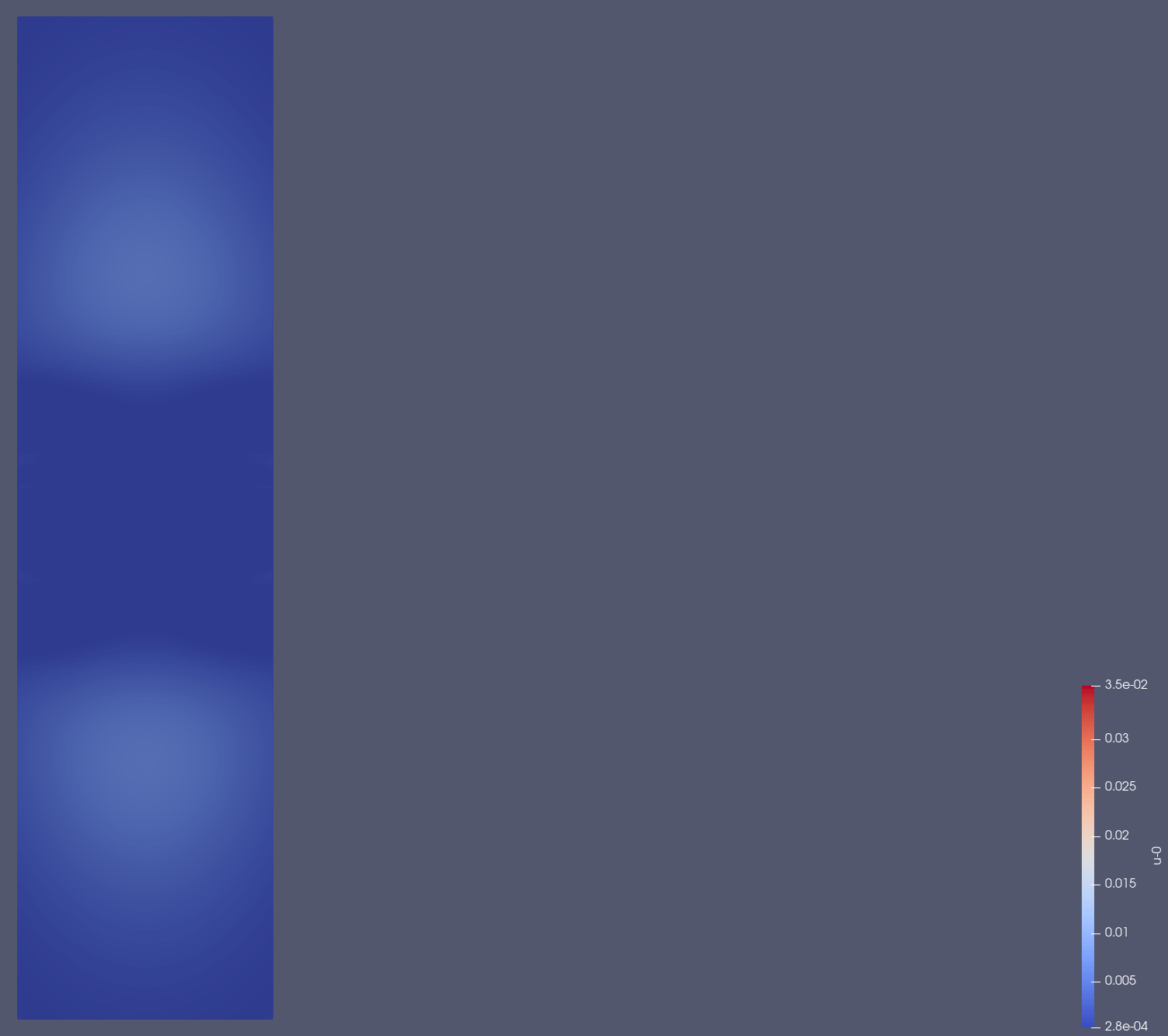}
\includegraphics[width=0.49\textwidth]{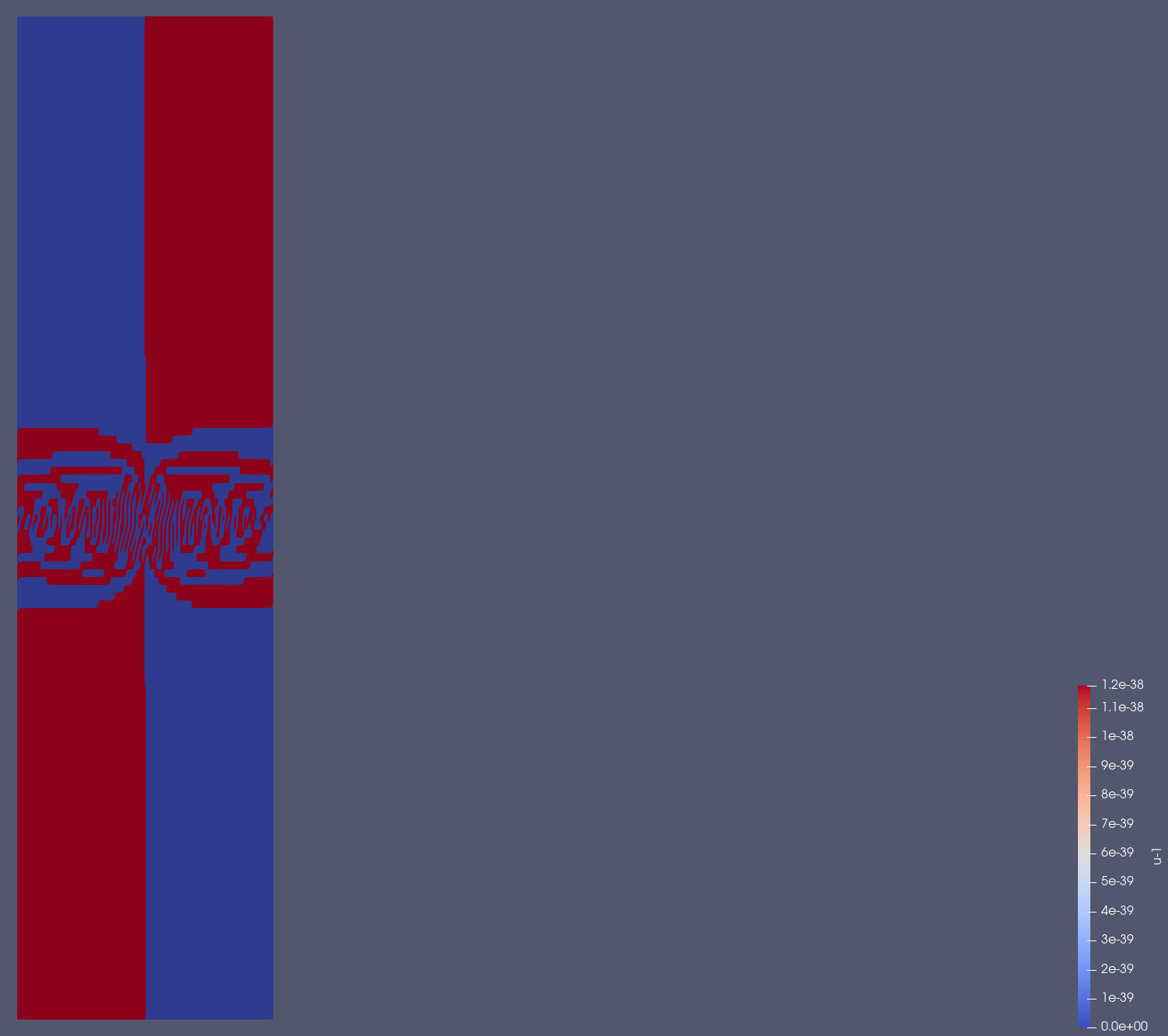}
\caption{Numerical solution of the real
(left) and imaginary (right) parts of the density matrix in convenient coordinates under a harmonic potential after an evolution time of $t=2.0$, solved by DG. Remark: Color scale differs between right picture and left one.}
\end{figure}

\begin{figure}[ht]
\centering
\includegraphics[width=0.49\textwidth]{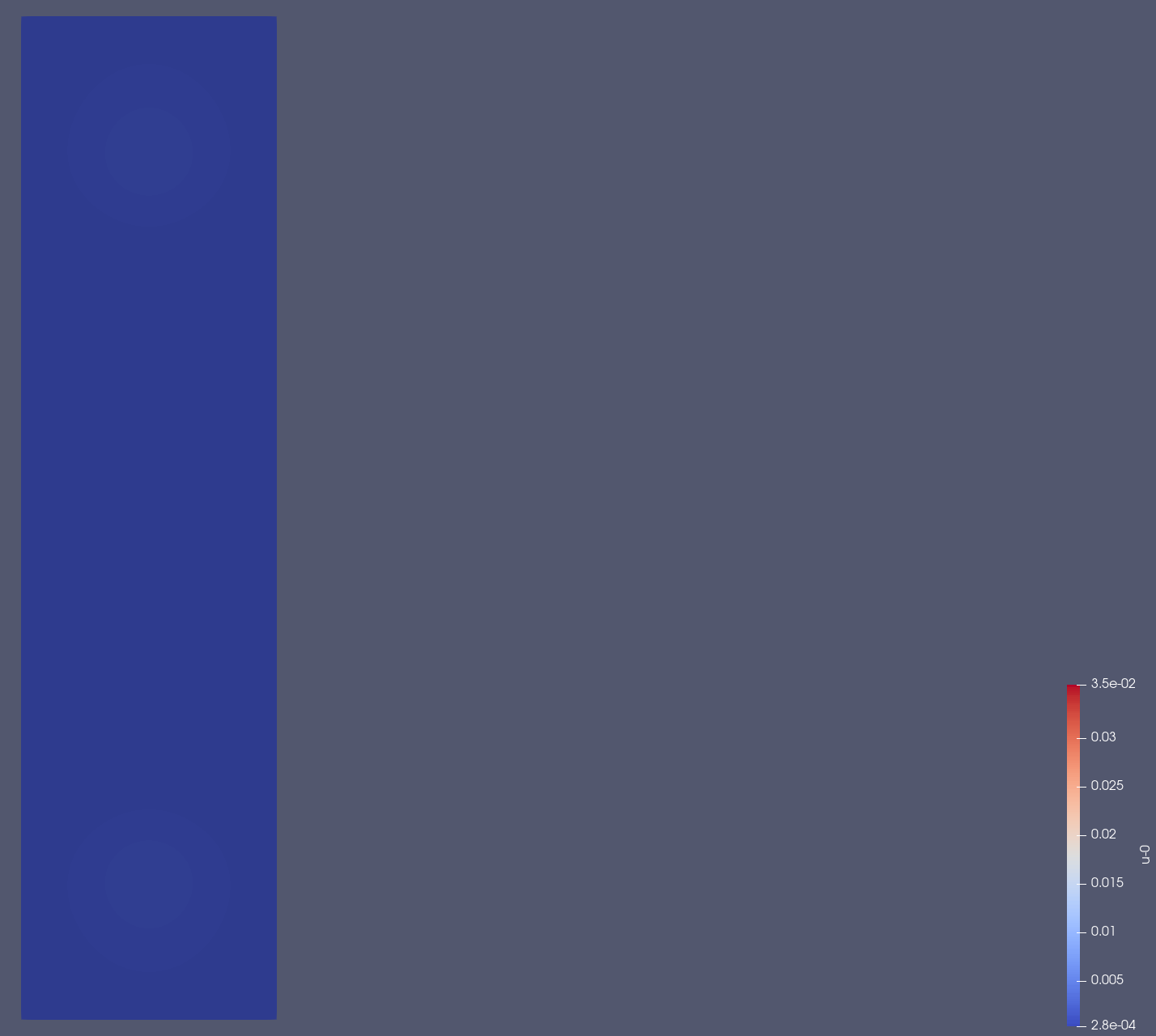}
\includegraphics[width=0.49\textwidth]{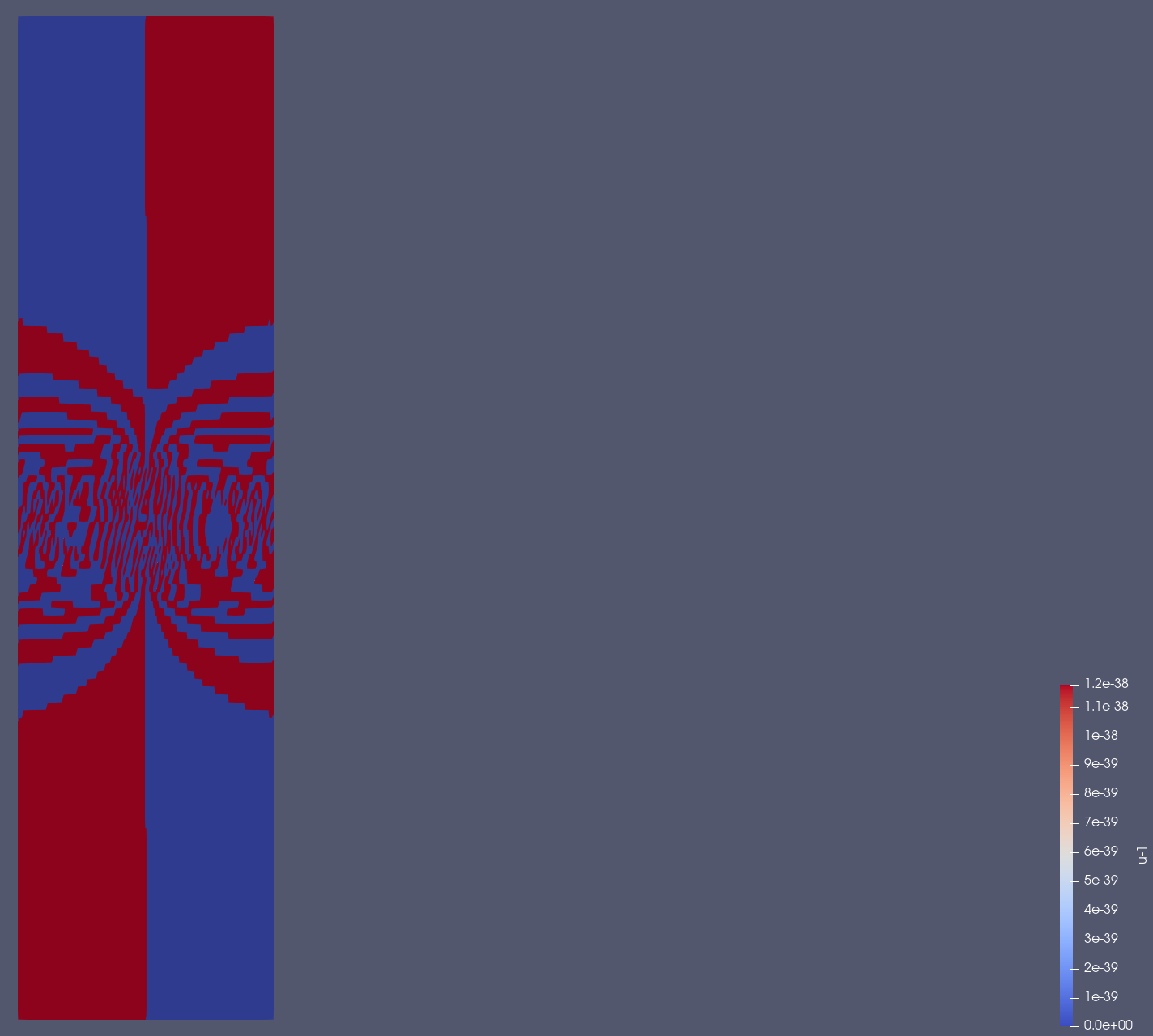}
\caption{Numerical solution of the real
(left) and imaginary (right) parts of the density matrix in convenient coordinates under a harmonic potential after an evolution time of $t=10.0$, solved by DG. Remark: Color scale differs between right picture and left one.}
\end{figure}

\begin{figure}[ht]
\centering
\includegraphics[width=0.49\textwidth]{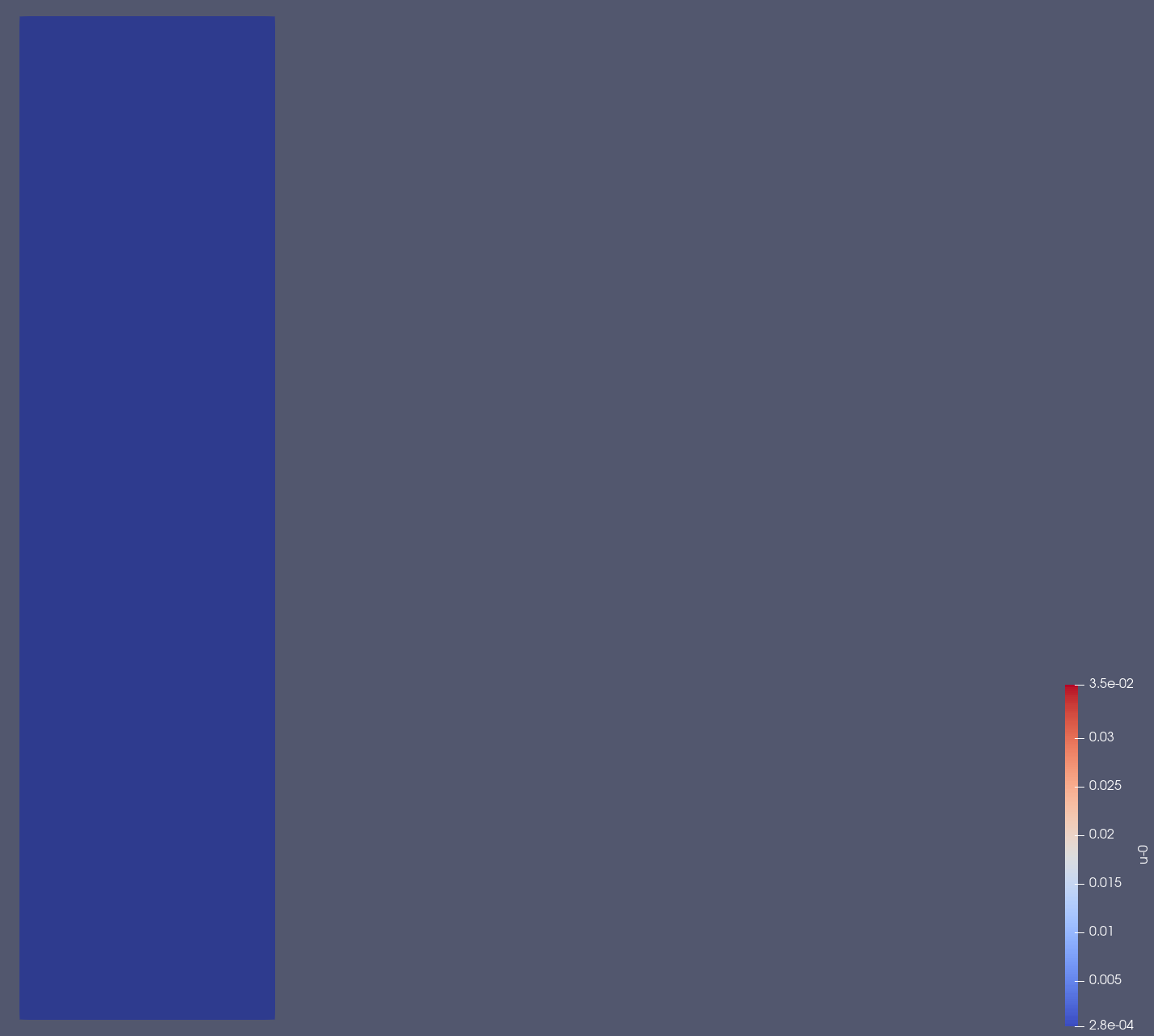}
\includegraphics[width=0.49\textwidth]{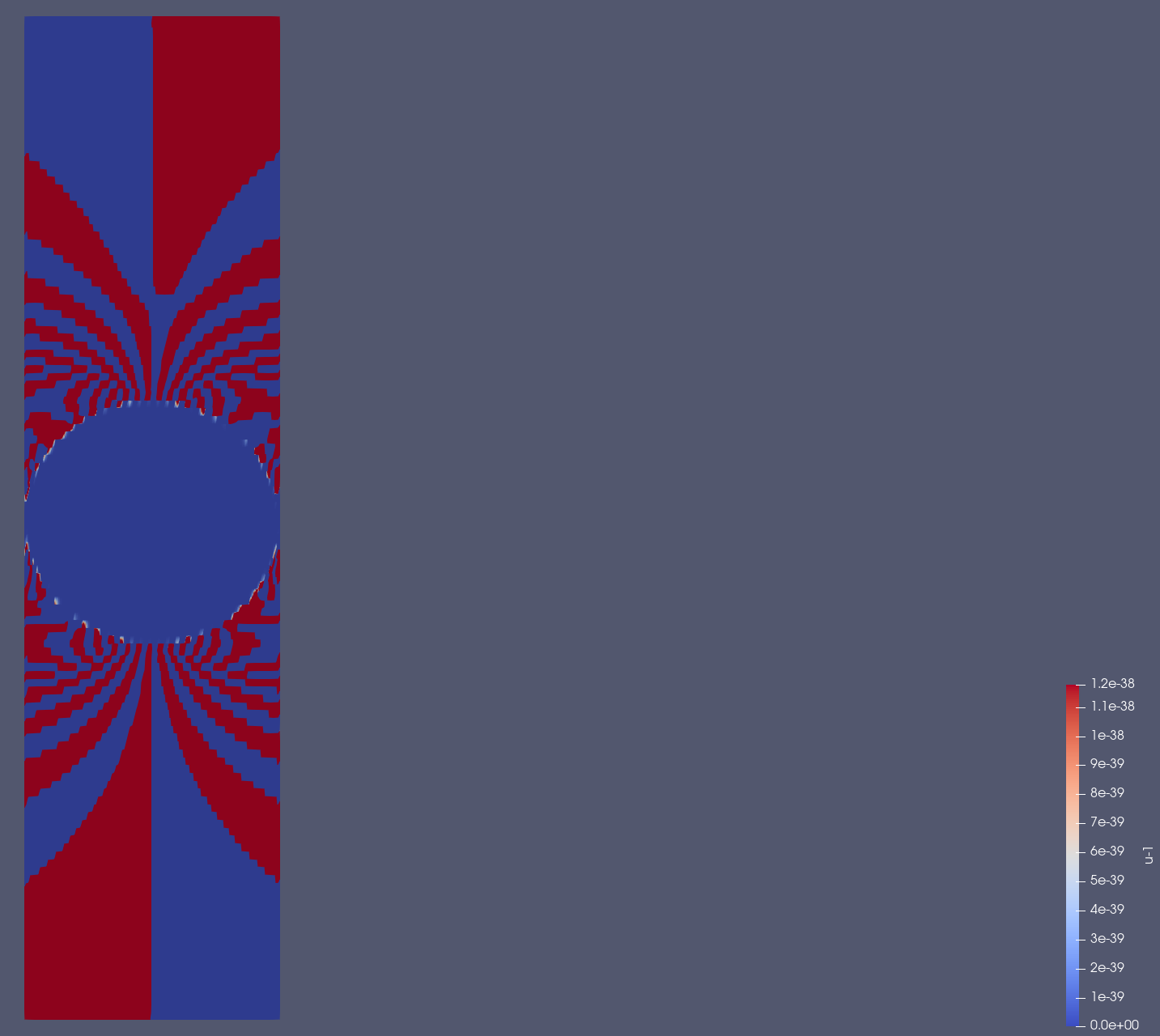}
\caption{Numerical steady state solution of the real
(left) and imaginary (right) parts of the density matrix in convenient coordinates under a harmonic potential after an evolution time of $t=50.0$, solved by DG. Remark: Color scale differs between right picture and left one.}
\end{figure}

\subsubsection{Convergence and Error Analysis for NIPG solutions at $t=2.0$}

The following table 
contains the $L_2$ error between the analytical steady state solution
$u_\infty = R_\infty+ iI_\infty$
and the numerical solution 
$u_h|_{t_f}=R_h|_{t_f}+iI_h|_{t_f}$
achieved after a time 
of $t_f=2.0$ (in normalized units), for both the real and imaginary components of the density matrix in convenient position coordinates.
This error is indicated for the different number of intervals in which each dimension is subdivided, with
 the same number of subdivisions 
 $N_x=N_{\eta}$
 in $x$ as in 
$\eta$. 

\begin{table}[ht!]
\centering
\begin{tabular}{| c| c |c | c|}
\hline
 $N_x=N=N_{\eta}$ & $||R_{\infty}-R_h|_{t_f}||_2$ & $||I_{\infty}-I_h|_{t_f}||_2$ & $||u_{\infty}-u_h|_{t_f}||_2$ \\
 \hline
 32 & 4.1992 & 1.1630 & 4.3573
 \\  
 \hline                
 64 & 2.6410 & 0.7913 & 2.7570 \\  
 \hline
 128 & 1.0793 & 0.3235 & 1.1267 \\   
 \hline
\end{tabular}
\caption{Comparison of the $L_2$ error in the real and imaginary components of the density matrix  between our numerical solution and the known analytical steady state for a harmonic oscillator benchmark problem, under a NIPG solver using a $V_h^1$ DG FE space, after an evolution time of $t_f=2.0$.}
\label{table:1}
\end{table}

The error roughly halves when refining by a factor of 2 the meshing in both $x$ and $\eta$, which is starting to indicate  an order of convergence of the method of the type
$\varepsilon=O(h^\kappa),\, \kappa=1$ (the numerical value that is obtained by the standard fit in the error analysis is $\kappa_{\mathrm{NSfit}}=0.9756$), as piece-wise linear polynomials (degree $\kappa=1$) 
have been used 
for our simulations.

For comparison, 
a table is presented as well 
where the $L_2$ error between the analytical form of the initial condition
$u_0 = R_0+ iI_0$
and its projection $u_h|_{t_0}=R_h|_{t_0}+iI_h|_{t_0}$ into the DG FE space
of piece-wise linear polynomials $V_h^1$
(where $t_0=0$) is indicated for the different number of intervals in which each dimension is subdivided, where again 
 $N_x=N=N_{\eta}$.
\begin{table}[ht!]
\centering
\begin{tabular}{| c| c |c |}
\hline
 $N_x=N=N_{\eta}$ & $||R_{0}-R_h|_{t_0}||_2=||u_{0}-u_h|_{t_0}||_2$ & $||I_{0}-I_h|_{t_0}||_2$ \\
 \hline
 32 & 0.0167 & 0.0
 \\  
 \hline                
 64 & 0.0042 & 0.0
 \\  
 \hline
 128 & 0.0010 & 0.0
 \\   
 \hline
\end{tabular}
\caption{Comparison of the $L_2$ projection error in the real and imaginary components of the density matrix in the initial condition between its analytical form and its numerical representation in a $V_h^1$ DG FE space.}
\label{table:2}
\end{table}
In this case, one can observe that the projection error behaves 
as in
$\varepsilon=O(h^{2\kappa}),\, \kappa=1$, using piece-wise linear polynomials. The actual fitted numerical value in the error analysis is $\kappa_{\mathrm{ICfit}}=1.0154$.

A more detailed analysis of the above-mentioned errors is  presented below, but now for the case when the number of intervals in $x$ and $\eta$ might differ, for both the projection $L_2$ error of the initial condition and the convergence error for the numerical solution versus the steady state after the $t_f=2.0$ time. 

\begin{table}[ht!]
\centering
\begin{tabular}{| c| c |c |c |}
\hline
 $||R_{0}-R_h|_{t_0}||_2$ & $N_{\eta}=32$  & $N_{\eta}=64$ & $N_{\eta}=128$ \\
 \hline
 $N_x=32$ & {\color{black}0.0167}  & 0.0062 & 0.0039
 \\  
 \hline                
 $N_x=64$ & 0.0150 &{\color{black} 0.0042} & 0.0016 \\  
 \hline
 $N_x=128$ & 0.0147 & 0.0038  &{\color{black} 0.0010}\\   
 \hline
\end{tabular}
\caption{Comparison of the $L_2$ projection error in the real component (for the imaginary part it is zero) of the density matrix in the initial condition between its analytical form and its numerical representation in a $V_h^1$ DG FE space.}
\end{table}

\begin{table}[ht!]
\centering
\begin{tabular}{| c| c |c |c |}
\hline
 $||R_{\infty}-R_h|_{t_f}||_2$ & $N_{\eta}=32$  & $N_{\eta}=64$ & $N_{\eta}=128$ \\
 \hline
 $N_x=32$ & {\color{black}4.1992} & 2.6420 & 1.0815
 \\  
 \hline                
 $N_x=64$ & 4.1986 & {\color{black}2.6410} & 1.0794 \\  
 \hline
 $N_x=128$ & 4.1993 &  2.6414 & {\color{black}1.0793}\\   
 \hline
\end{tabular}
\caption{Comparison of the $L_2$ error in the real component of the density matrix between our numerical solution and the known analytical steady state for a harmonic oscillator benchmark problem, under a NIPG solver using a $V_h^1$ DG FE space, after an evolution time of $t_f=2.0$.}
\end{table}

\begin{table}[ht!]
\centering
\begin{tabular}{| c| c |c |c |}
\hline
 $||I_{\infty}-I_h|_{t_f}||_2$ & $N_{\eta}=32$  & $N_{\eta}=64$ & $N_{\eta}=128$ \\
 \hline
 $N_x=32$ & {\color{black} 1.1630} & 0.7872 & 0.3210
 \\  
 \hline                
 $N_x=64$ & 1.1662 &{\color{black} 0.7913} & 0.3227 \\  
 \hline
 $N_x=128$ & 1.1678 & 0.7932  & {\color{black} 0.3235}\\   
 \hline
\end{tabular}
\caption{Comparison of the $L_2$ error in the  imaginary component of the density matrix between our numerical solution and the known analytical steady state for a harmonic oscillator benchmark problem, under a NIPG solver using a $V_h^1$ DG FE space, after an evolution time of $t_f=2.0$.}
\end{table}

The behavior of the $L_2$ error regarding the numerical  solution after an evolution time of $t_f=2.0$ can be explained by understanding that our phenomena is of a convective-diffusive type, where the convective part might dominate over the diffusive one, and since the transport is mostly vertical, the refinement in $\eta$ is the important one regarding error behavior due to mesh discretization, as opposed to the $x$-refinement which doesn't have as much an effect in changing the aforementioned error.

\subsubsection{Convergence and Error Analysis for NIPG solutions at $t=50.0$}

The table presented below
contains the $L_2$ error between the analytical steady state solution
$u_\infty = R_\infty+ iI_\infty$
and the numerical solution 
$u_h|_{t_f}=R_h|_{t_f}+iI_h|_{t_f}$
achieved after a time of 
$t_f=50.0$ (where the numerical solution is close to the analytical steady state after that evolution time), for both the real and imaginary components of the density matrix in convenient position coordinates.
This error is indicated for the different number of intervals in which each dimension is subdivided, with
 the same number of subdivisions 
 $N_x=N_{\eta}$
 in $x$ as in 
$\eta$. 

\begin{table}[ht!]
\centering
\begin{tabular}{| c| c |c | c|}
\hline
 $N_x=N=N_{\eta}$ & $||R_{\infty}-R_h|_{t_f}||_2$ & $||I_{\infty}-I_h|_{t_f}||_2$ & $||u_{\infty}-u_h|_{t_f}||_2$ \\
 \hline
 2 & 7.2349 & 0.0283  & 7.2350 \\   
  \hline
 4 & 5.3768 & 0.5212 & 5.4020 \\   
  \hline
 8 & 3.9771 & 0.8732 & 4.0718 \\   
  \hline
 16 & 2.2221 & 0.5987 & 2.3014 \\   
  \hline
 32 & 0.9311  & 0.2107 & 0.9546 \\   
  \hline
 64 & 0.4604 & 0.0261 & 0.4612 \\   
  \hline
   128 & 0.3352 & 0.0246 & 0.3361 \\   
  \hline
  \end{tabular}
\caption{Comparison of the $L_2$ error in the real and imaginary components of the density matrix in the steady state between our numerical solution and the known analytical one for a harmonic oscillator benchmark problem, under a NIPG solver using a $V_h^1$ DG FE space, after an evolution time of $t_f=50.0$.}
\label{table:3}
\end{table}

The numerical value for the error
convergence rate
obtained by the standard fit 
for $||u_{\infty}-u_h|_{t_f}||_2$
is
of the type
$\varepsilon=O(h^\kappa),\, \kappa_{\mathrm{NSfit}}=0.9517$.
Let's remember that 
piece-wise linear polynomials (degree $\kappa=1$) 
have been used 
for our simulations.

For comparison, 
a table is presented as well 
where the $L_2$ error between the analytical form of the initial condition
$u_0 = R_0+ iI_0$
and its projection $u_h|_{t_0}=R_h|_{t_0}+iI_h|_{t_0}$ into the DG FE space
of piece-wise linear polynomials $V_h^1$
(where $t_0=0$) is indicated for the different number of intervals in which each dimension is subdivided, where again 
 $N_x=N=N_{\eta}$.
\begin{table}[ht!]
\centering
\begin{tabular}{| c| c |c |}
\hline
 $N_x=N=N_{\eta}$ & $||R_{0}-R_h|_{t_0}||_2=||u_{0}-u_h|_{t_0}||_2$ & $||I_{0}-I_h|_{t_0}||_2$ 
 \\
 \hline
 2 & 1.4274 & 0.0
 \\
 \hline
 4 & 0.8582 & 0.0
 \\
 \hline
 8 & 0.2599  & 0.0
 \\
  \hline
 16 & 0.0664 & 0.0
 \\
 \hline
 32 & 0.0167 
 & 0.0
 \\  
 \hline                
 64 & 0.0042 & 0.0
 \\  
 \hline
 128 & 0.0010 & 0.0
 \\   
 \hline
\end{tabular}
\caption{Comparison of the $L_2$ projection error in the real and imaginary components of the density matrix in the initial condition between its analytical form and its numerical representation in a $V_h^1$ DG FE space.}
\label{table:4}
\end{table}
In this case, one can observe that the projection error behaves 
as in
$\varepsilon=O(h^{2\kappa}),\, \kappa=1$, using piece-wise linear polynomials (the actual fitted numerical value in the error analysis is $\kappa_{\mathrm{ICfit}}=0.9952$).

A more detailed analysis of the above-mentioned errors is  presented below, but now for the case when the number of intervals in $x$ and $\eta$ might differ, for both the projection $L_2$ error of the initial condition and the convergence error for the numerical solution of the steady state after our evolution time of $t_f=50$. 

\begin{table}[ht!]
\centering
\begin{tabular}{| c| c |c |c |c |c |}
\hline
 $||R_{0}-R_h|_{t_0}||_2$ & $N_{\eta}=8$ &  $N_{\eta}=16$ & $N_{\eta}=32$  & $N_{\eta}=64$ & $N_{\eta}=128$  \\
 \hline
 $N_x=8$ & {\color{black} 0.2599}  & 0.0982  & 0.0624 & 0.0550 & 0.0534
 \\  
 \hline                
 $N_x=16$ & 0.2348 &{\color{black} 0.0664} & 0.0248 & 0.0157 & 0.0139 \\  
 \hline
 $N_x=32$ & 0.2291 & 0.0599  &{\color{black} 0.0167} & 0.0062 & 0.0039  \\   
 \hline
  $N_x=64$ & 0.2277 & 0.0584   & 0.0150  & {\color{black} 0.0042} & 0.0015 \\   
 \hline
 $N_x=128$ & 0.2273 & 0.0580  & 0.0146  & 0.0037 & {\color{black} 0.0010}\\   
 \hline
\end{tabular}
\caption{Comparison of the $L_2$ projection error in the real component (for the imaginary part it is zero) of the density matrix in the initial condition between its analytical form and its numerical representation in a $V_h^1$ DG FE space.}
\end{table}

\begin{table}[ht!]
\centering
\begin{tabular}{| c| c |c |c |c |c |}
\hline
 $||R_{\infty}-R_h|_{t_\infty}||_2$ & $N_{\eta}=8$ &  $N_{\eta}=16$ & $N_{\eta}=32$  & $N_{\eta}=64$ & $N_{\eta}=128$  \\
 \hline
 $N_x=8$ & {\color{black} 3.9771}  & 2.2802 & 0.9567 & 0.5039 & 0.3632
 \\  
 \hline                
 $N_x=16$ & 3.8911  &{\color{black} 2.2221} & 0.8727 & 0.4183 & 0.3004 \\  
 \hline
 $N_x=32$ & 3.9393 & 2.2722  &{\color{black} 0.9311} & 0.4764 & 0.3447 \\   
 \hline
  $N_x=64$ & 3.9233 & 2.2603 & 0.9151 & {\color{black} 0.4604} & 0.3359 \\   
 \hline
 $N_x=128$ & 3.9222 & 2.2591  & 0.9131 & 0.4583 & {\color{black}0.3352 }\\   
 \hline
\end{tabular}
\caption{Comparison of the $L_2$ error in the real component of the density matrix in the steady state between our numerical solution and the known analytical one for a harmonic oscillator benchmark problem, under a NIPG solver using a $V_h^1$ DG FE space, after an evolution time of $t_f=50.0$.}
\end{table}

\begin{table}[ht!]
\centering
\begin{tabular}{| c| c |c |c |c |c |}
\hline
 $||I_{\infty}-I_h|_{t_\infty}||_2$ & $N_{\eta}=8$ &  $N_{\eta}=16$ & $N_{\eta}=32$  & $N_{\eta}=64$ & $N_{\eta}=128$  \\
 \hline
 $N_x=8$ & {\color{black} 0.8732}  & 0.5843  & 0.2038 & 0.0252 & 0.0002
 \\  
 \hline                
 $N_x=16$ & 0.8865 &{\color{black} 0.5987} & 0.2085 & 0.0258  & 0.0103 \\  
 \hline
 $N_x=32$ & 0.8947 &  0.6056 &{\color{black}0.2107 } & 0.0260  & 0.0235 \\   
 \hline
  $N_x=64$ & 0.8993 & 0.6092 & 0.2117 & {\color{black}0.0261 } & 0.0248\\   
 \hline
 $N_x=128$ & 0.9016  &  0.6110 & 0.2122 & 0.0262 & {\color{black} 0.0246}\\   
 \hline
\end{tabular}
\caption{Comparison of the $L_2$ error in the  imaginary component of the density matrix in the steady state between our numerical solution and the known analytical one for a harmonic oscillator benchmark problem, under a NIPG solver using a $V_h^1$ DG FE space, after an evolution time of $t_f=50.0$.}
\end{table}

The behavior of the $L_2$ error regarding the numerical steady state solution can be explained again by understanding that our phenomena is of a convective-diffusive type, but where the convective part dominates over the diffusive one. Since the transport is mostly vertical, the refinement in $\eta$ is the important one regarding error behavior due to mesh discretization, as opposed to the $x$-refinement, which doesn't seem to have as much an effect again in changing the aforementioned error.

\subsection{Comparing Homogeneous vs. Steady-state BC for Benchmark Harmonic Case}

Because applying as boundary conditions the analytical solution for the steady state is only helpful for the harmonic benchmark problem in which that steady state is known, we compare the performance of this BC case against the case when homogeneous BC are applied in the particular case of the benchmark harmonic potential. This is done to see the difference in performance against the more generally common case of applying homogeneous BC, as these are the most reasonable boundary conditions when the domain is big enough: the diffusive transport problem will not reach those regions and therefore the solution will be analytically/numerically zero over that boundary.

We present below a plot comparing the $L_2$-distance between the analytically known steady state solution to the harmonic benchmark problem and the numerical solutions obtained with either homogeneous BC or steady state BC formulas at a time of $t=50$ in the chosen units, for different mesh refinements.

\begin{figure}[ht]
\centering
\includegraphics[width=1.0\textwidth]{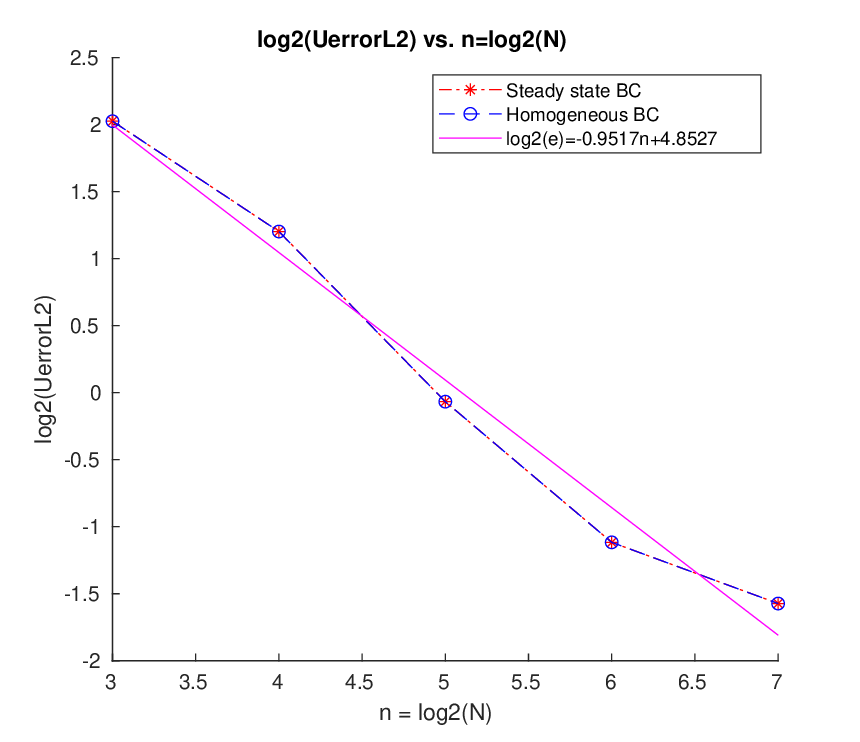}
\caption{Plot of the base 2 logarithm of the error between the numerical solution at the long time of $t=50$ and the analytical solution to the steady state versus the base 2 logarithm of the number of intervals in which both axis are refined, for the benchmark problem of a harmonic potential. Both cases of homogeneous BC and steady state BC formulas are presented, as well as the line that  fits both respective points. It is assumed the error behaves as $\epsilon\approx Ch^k$, where $k$ is the polynomial degree used. The slope of the fitted line is $k=0.9517\approx 1$, as expected when using linear polynomials in our Finite Element methodology.}
\label{fig:log2eVSn}
\end{figure}

Figure \label{fig:log2eVSn} 
clearly shows that there is no difference in the output results between applying homogeneous BC versus the steady-state analytical solution as BC at the long time of $t=50$, presumably because the Gaussian related to the latter has decayed long enough at the boundary so that the respective BC values are numerically zero. 
The $L_2$-error values are the same for both homogeneous BC and steady state BC for different mesh refinements at $t=50$, and therefore have the same line fitting their data points in the convergence analysis presented. 
The line fitted to the points respective to both BC has a negative slope close in absolute value to $k=1$, which is expected from using linear polynomials in our Finite Element Method. This is because it is expected that the error will behave as $\varepsilon=C h^k$, with $k$ the degree of the polynomial used in our Finite Element methodology. 
The moral of the story under this analysis is that our code is behaving as expected under mesh refinements, and that we can confidently use homogeneous BC for more general problems different to the harmonic benchmark case at high mesh refinements and long times. 
We will proceed then to study the case of a linear potential as a second problem after the benchmark.

\subsection{Linear Potential}

Our second problem of interest is the case of a linear potential
of the specific form $V(x)=x$. Up to our knowledge, there is no analytical solution for the steady state of this problem, which makes prescient the need of a computational solution for this problem. Our DG solver can help alleviate this issue by providing this computational solution, having previously benchmarked the code under the case of a harmonic potential for which the steady state analytical solution is known. 

We show the results of our simulations for the time evolution of WFP under a linear potential where the evolution time is $t=50$ in the chosen units, where the initial condition is the harmonic groundstate, and the boundary conditions are chosen to be homogeneous. Our plots represent the status of the real and imaginary components of our transformed density matrix in the chosen position coordinates at the times of $t=2$, $t=10$, and $t=50$ as the final time of the simulation for the last case.

\begin{figure}[ht]
\centering
\includegraphics[width=0.49\textwidth]{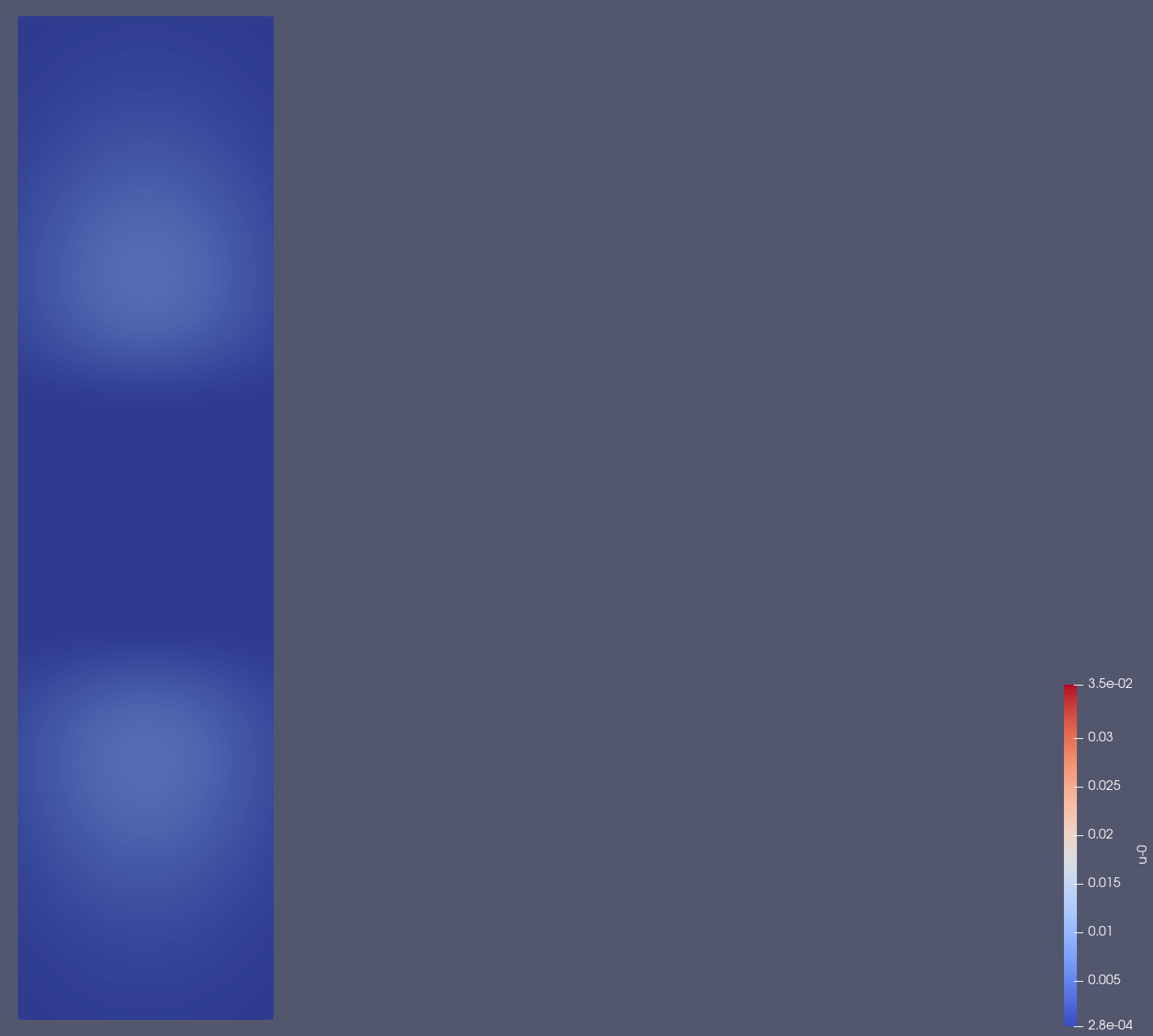}
\includegraphics[width=0.49\textwidth]{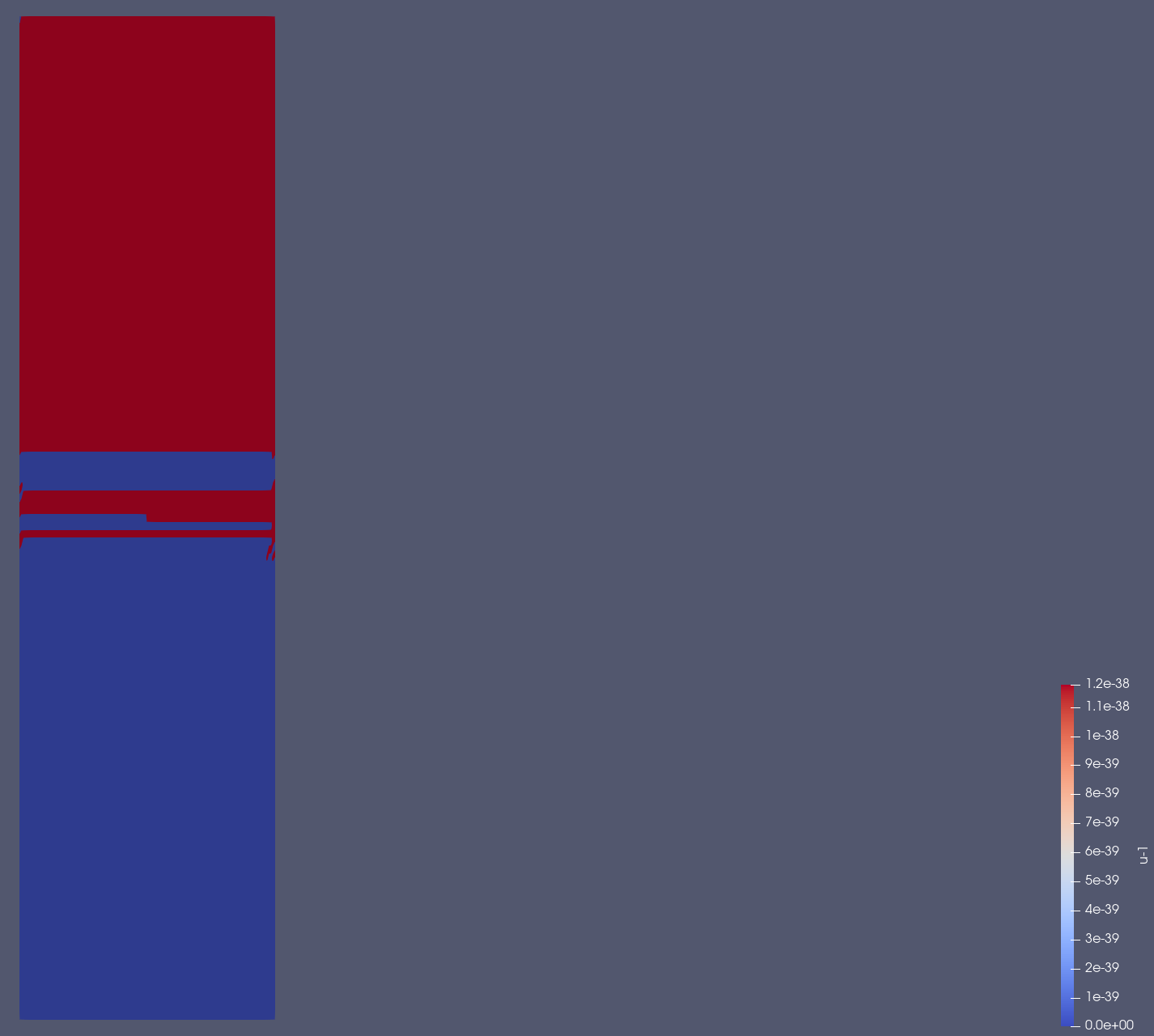}
\caption{Numerical solution of the real
(left) and imaginary (right) parts of the density matrix in convenient coordinates under a linear potential after an evolution time of $t=2.0$, solved by DG. Remark: Color scale differs between right picture and left one.}
\end{figure}

\begin{figure}[ht]
\centering
\includegraphics[width=0.49\textwidth]{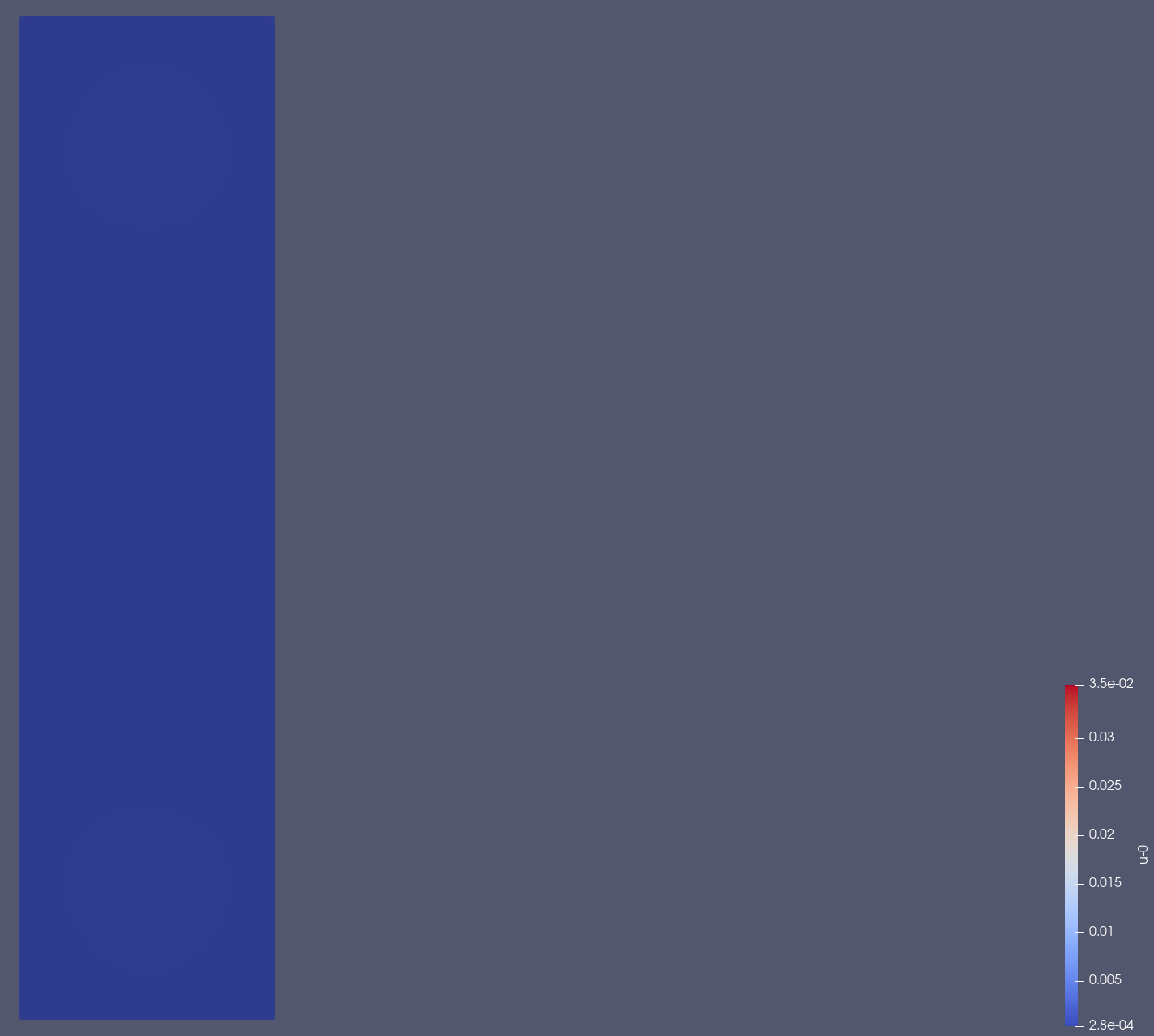}
\includegraphics[width=0.49\textwidth]{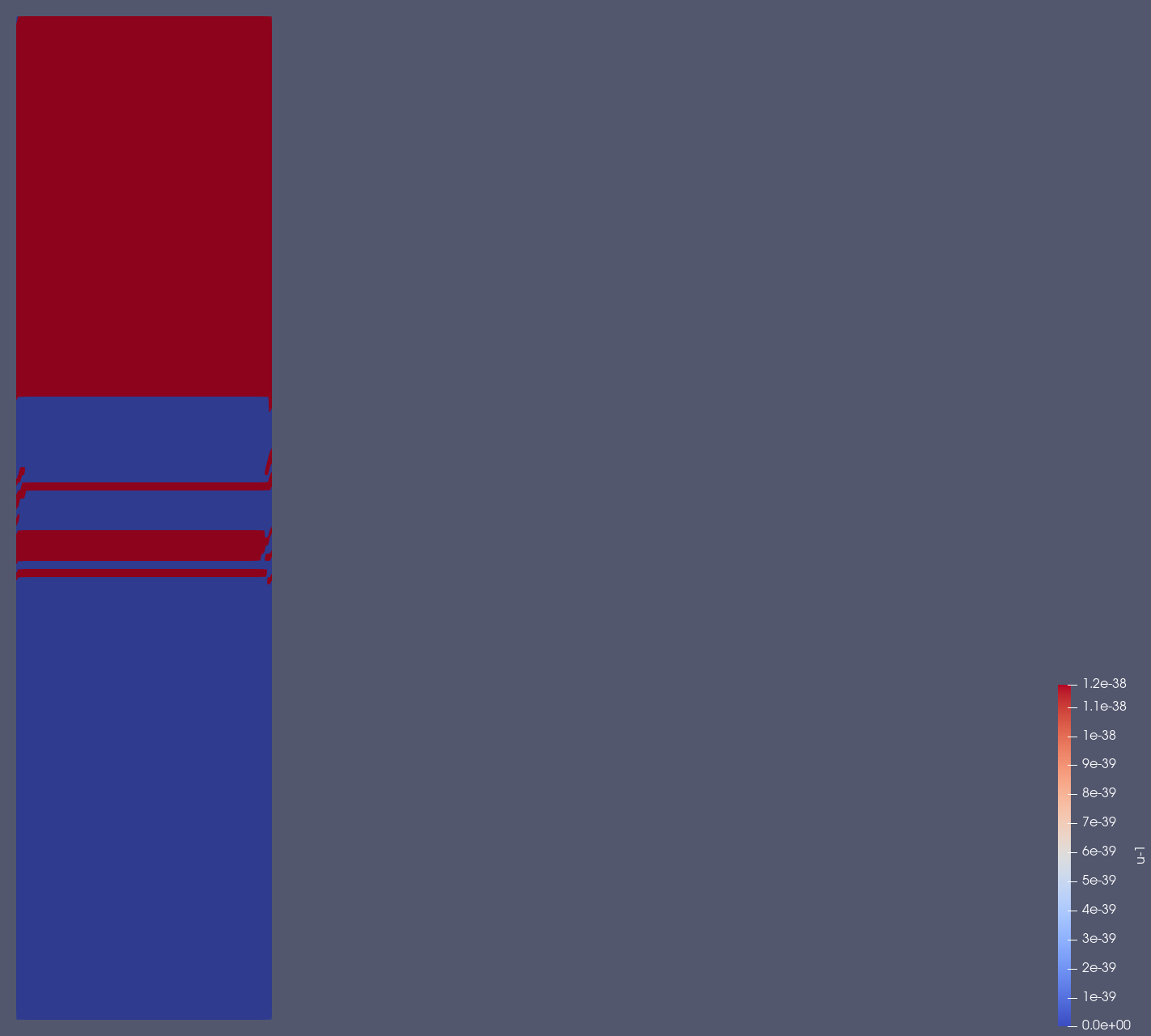}
\caption{Numerical solution of the real
(left) and imaginary (right) parts of the density matrix in convenient coordinates under a linear potential after an evolution time of $t=10.0$, solved by DG. Remark: Color scale differs between right picture and left one.}
\end{figure}

\begin{figure}[ht]
\centering
\includegraphics[width=0.49\textwidth]{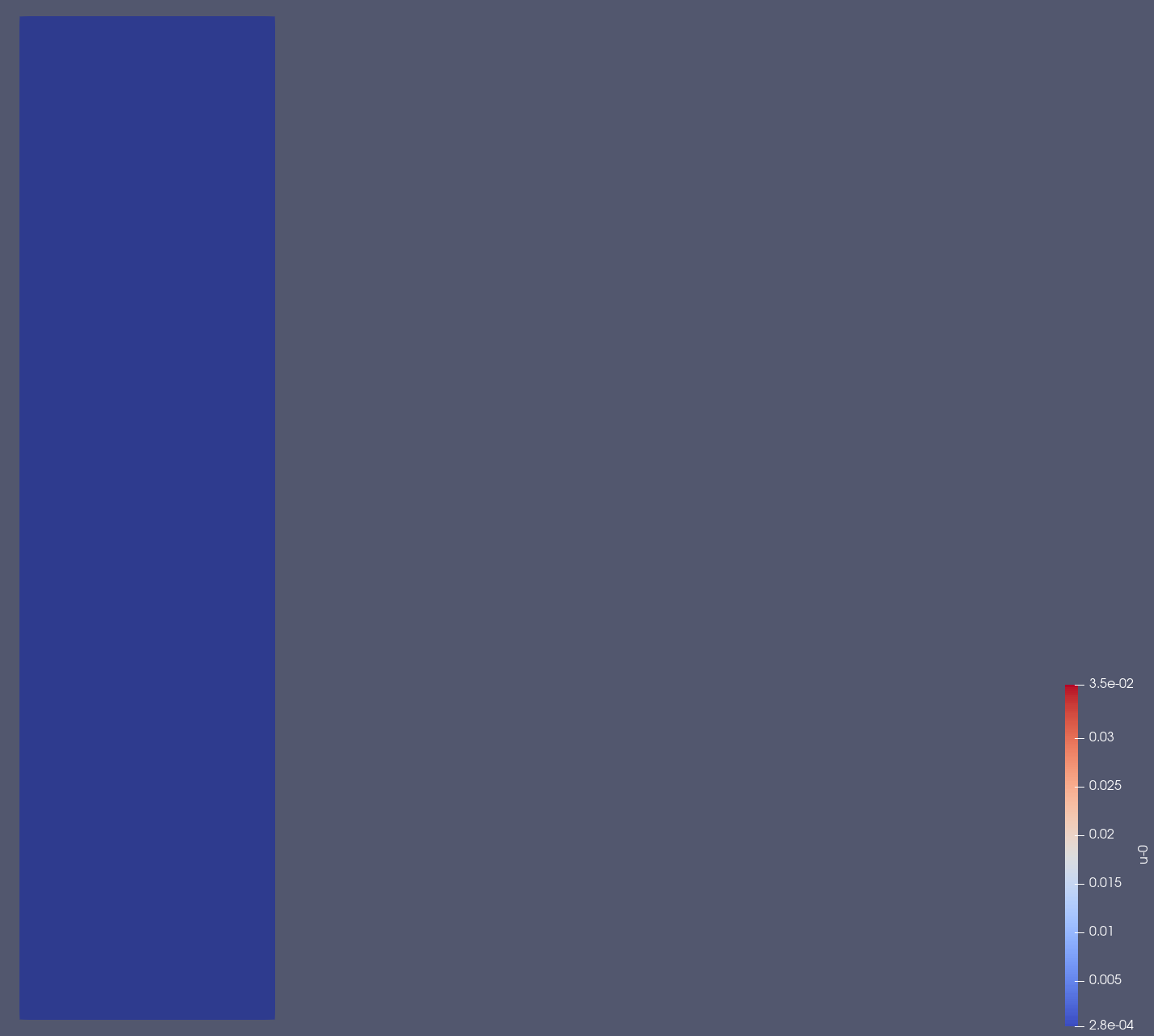}
\includegraphics[width=0.49\textwidth]{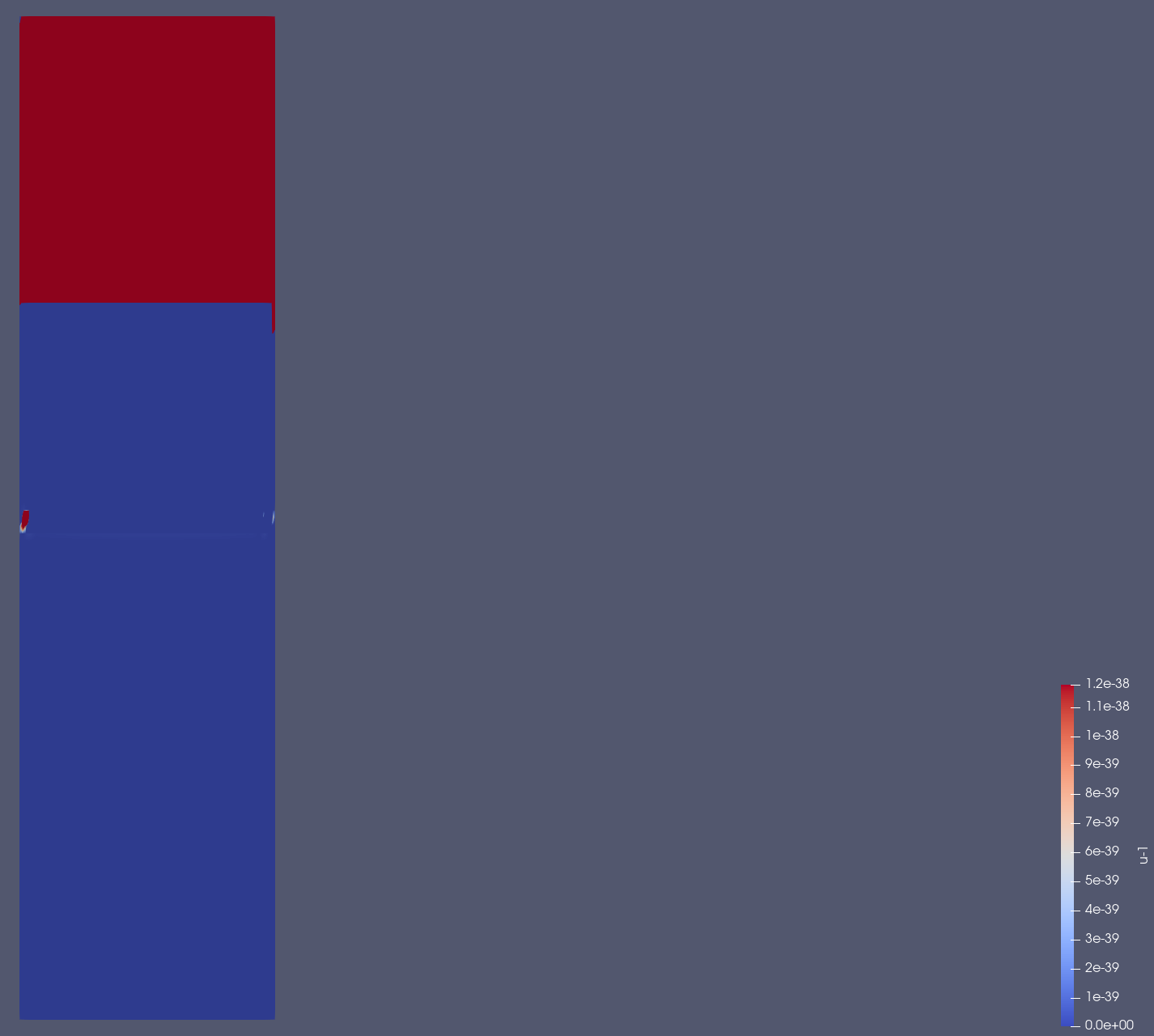}
\caption{Numerical steady state solution of the real
(left) and imaginary (right) parts of the density matrix in convenient coordinates under a linear potential after an evolution time of $t=50.0$, solved by DG. Remark: Color scale differs between right picture and left one.}
\end{figure}

\subsection{Quartic Potential}

We now focus on the case of a quartic potential. 
The motivation to do so is the fact that it is of higher degree than the quadratic potential benchmark, yet $V(x)=x^4$ is qualitatively similar to $V(x)=x^2$ regarding the global minimum and potential well behavior. The computational solution for WFP in the case of a quartic potential is of natural interest given that there is no analytical knowledge of the steady state solution for this case. Therefore the only way to get to know qualitatively and quantitatively the steady state solution for WFP with a quartic potential is through computational simulations. 

We show below the results of our simulations for the time evolution of WFP under a quartic potential. The evolution time is $t=50$ in the chosen units. The initial condition is the harmonic groundstate. The boundary conditions are chosen to be homogeneous. Our plots represent the status of the real and imaginary components of our transformed density matrix in the chosen position coordinates at the times of $t=2$, $t=10$, and $t=50$ respectively, taking the later as the numerical solution for the steady state. 

\begin{figure}[ht]
\centering
\includegraphics[width=0.49\textwidth]{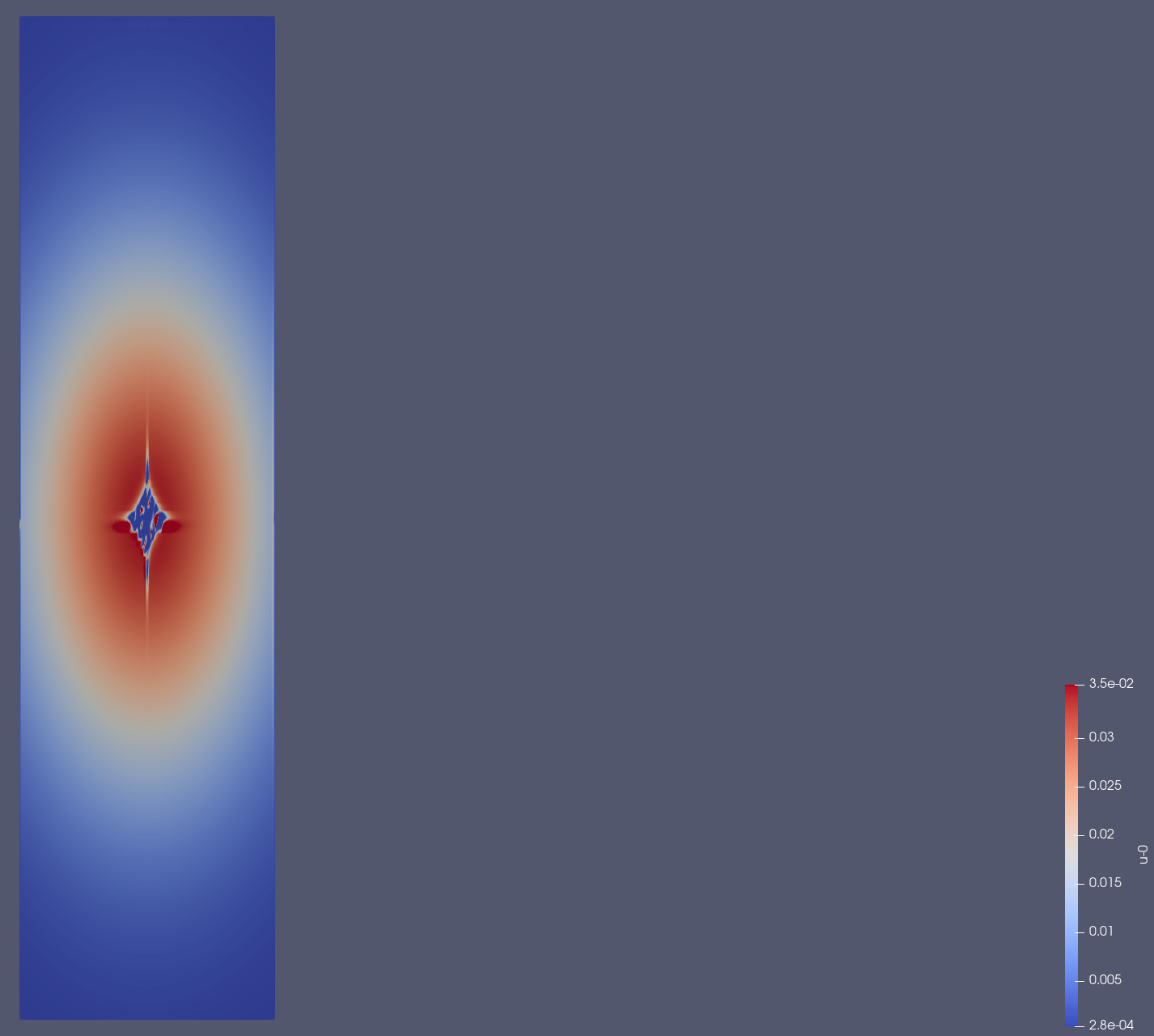}
\includegraphics[width=0.49\textwidth]{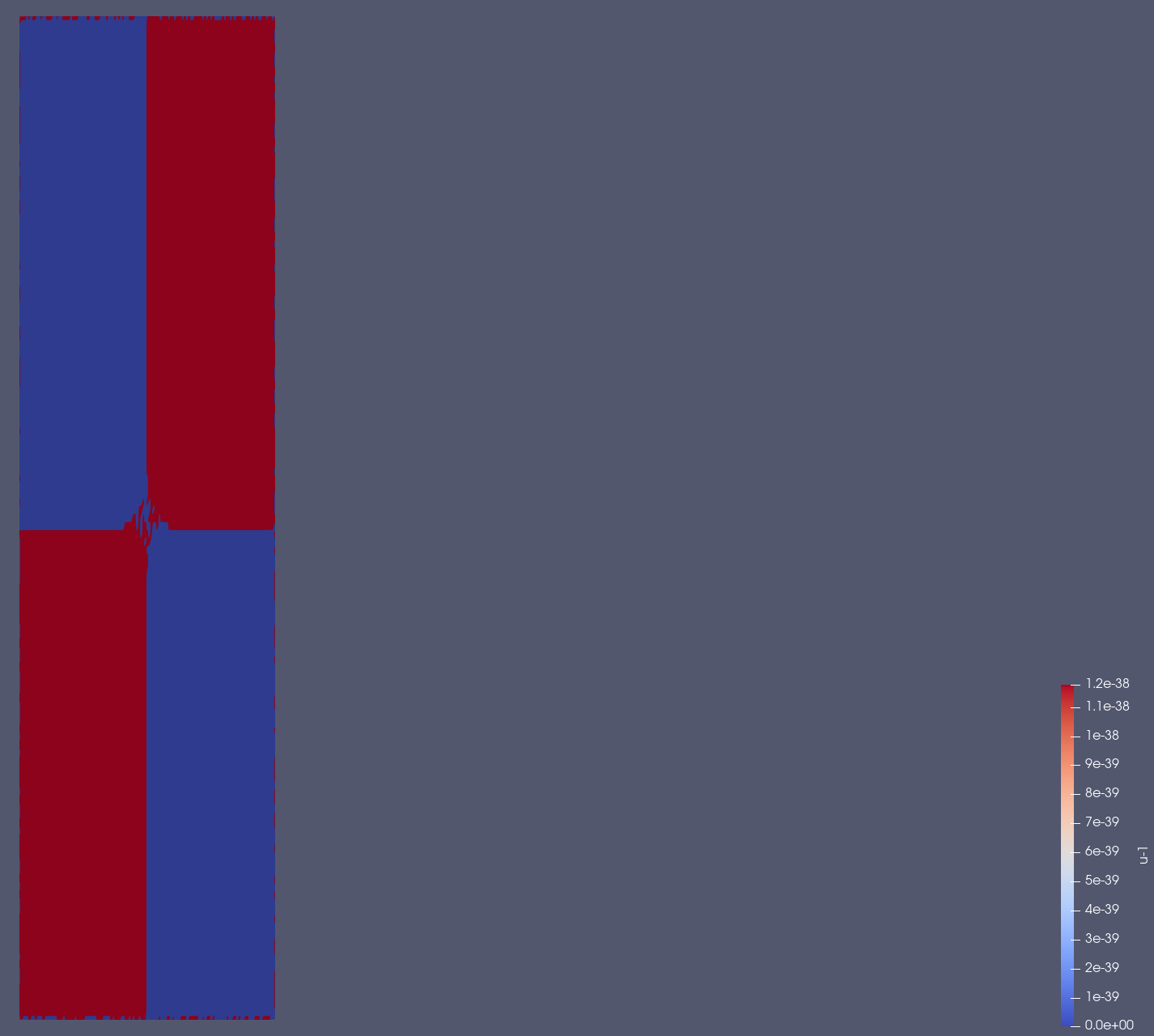}
\caption{Numerical solution of the real
(left) and imaginary (right) parts of the density matrix in convenient coordinates under a quartic potential after an evolution time of $t=2.0$, solved by DG. Remark: Color scale differs between right picture and left one.}
\end{figure}

\begin{figure}[ht]
\centering
\includegraphics[width=0.49\textwidth]{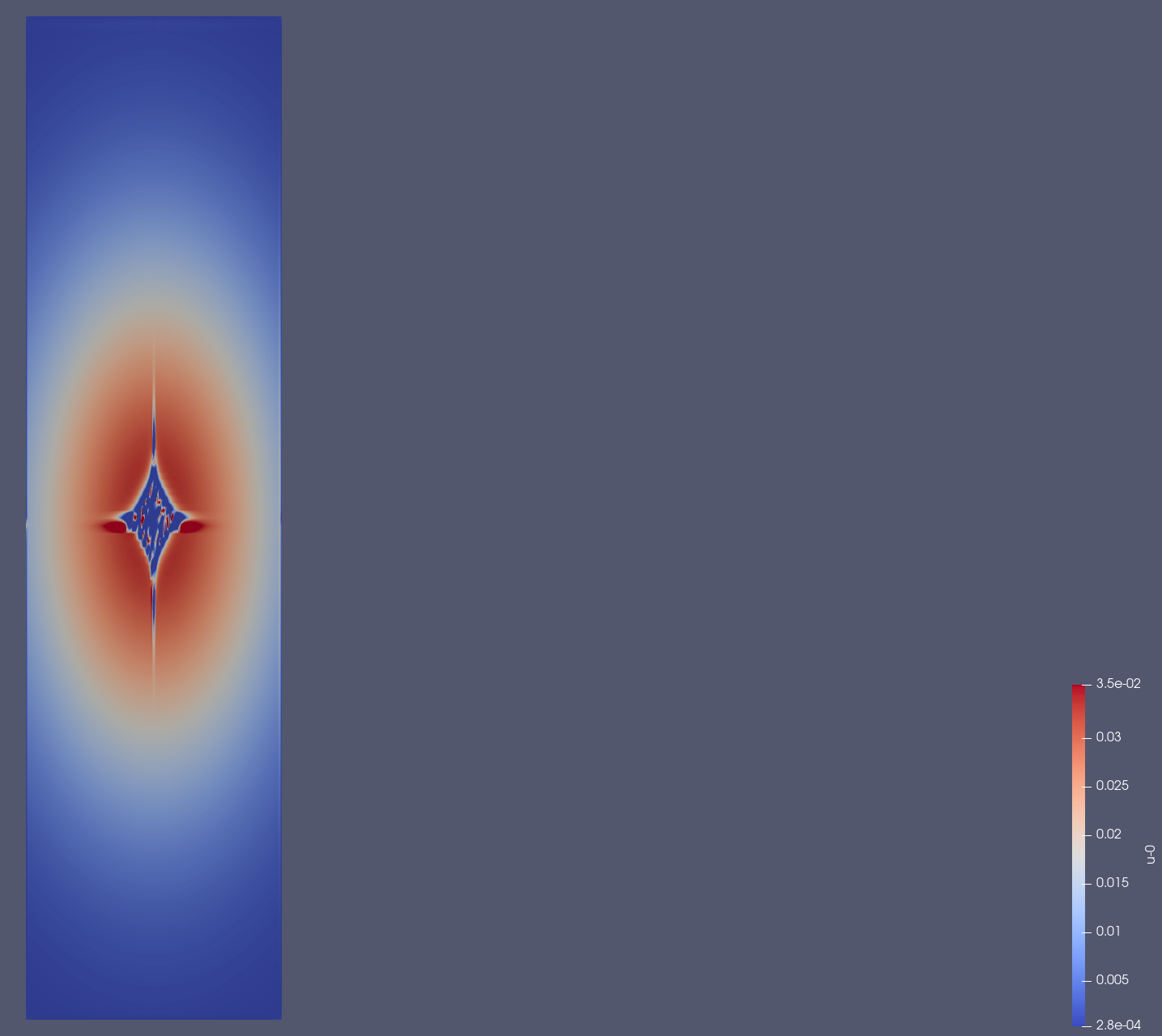}
\includegraphics[width=0.49\textwidth]{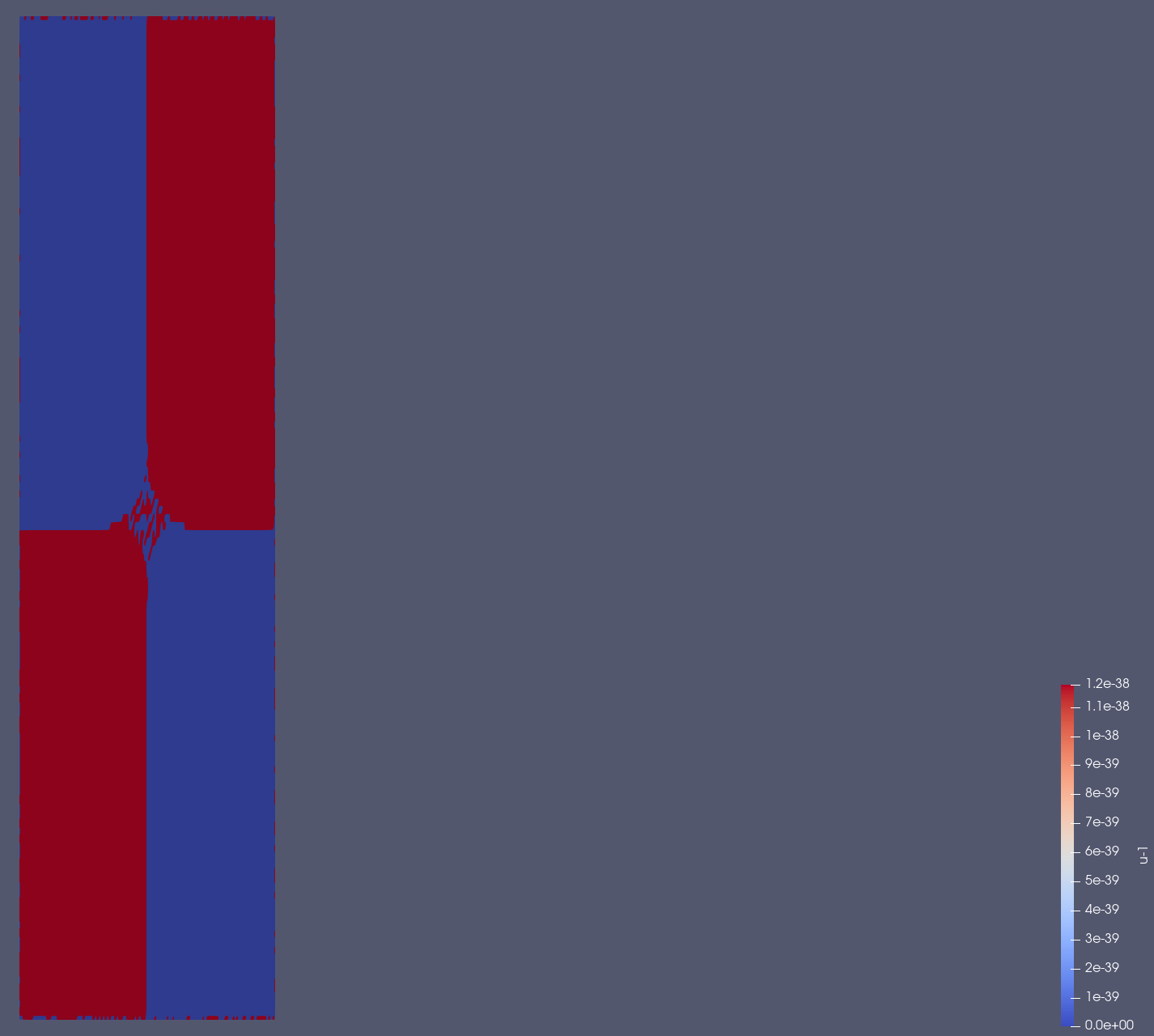}
\caption{Numerical solution of the real
(left) and imaginary (right) parts of the density matrix in convenient coordinates under a quartic potential after an evolution time of $t=10.0$, solved by DG. Remark: Color scale differs between right picture and left one.}
\end{figure}

\begin{figure}[ht]
\centering
\includegraphics[width=0.49\textwidth]{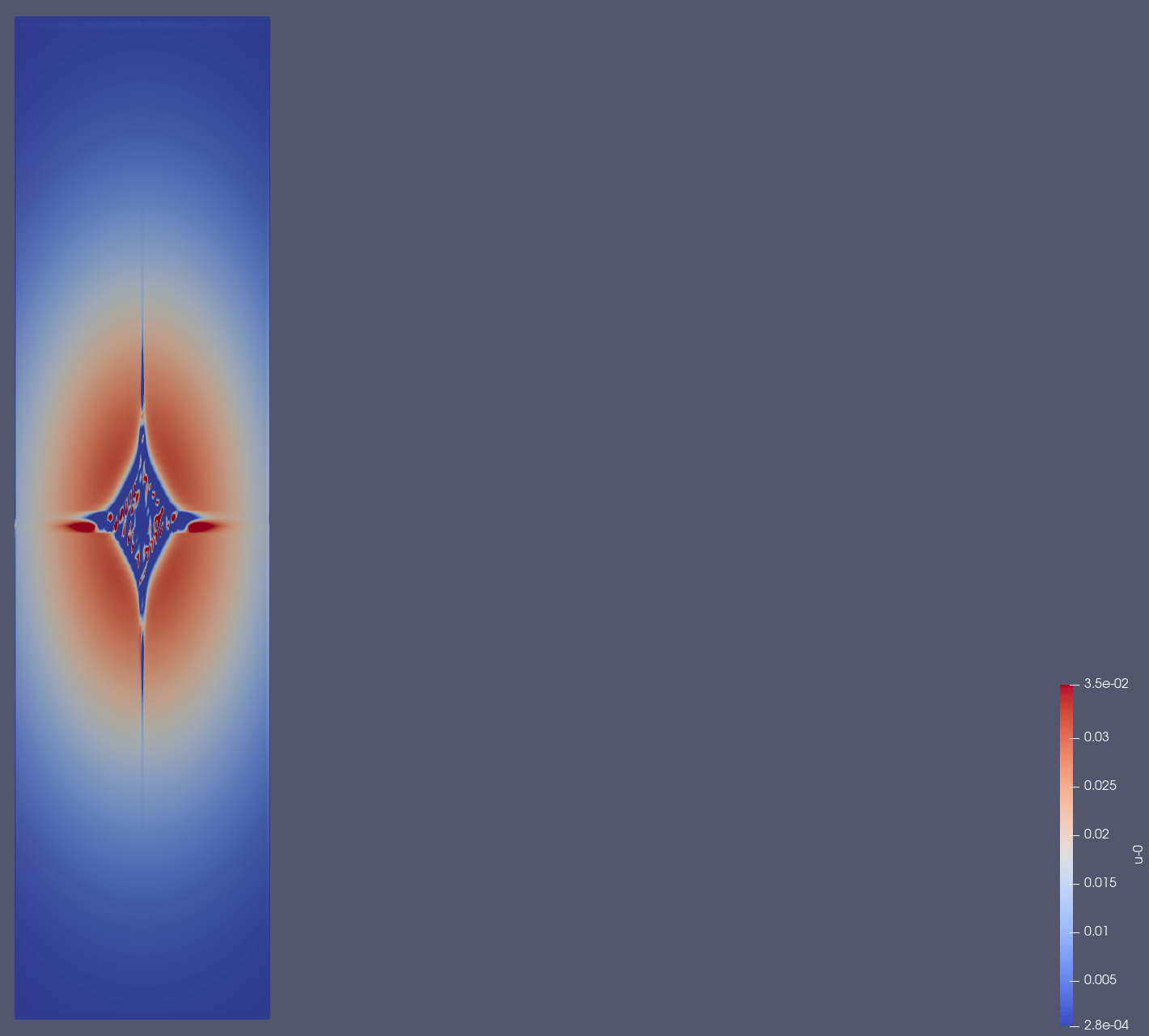}
\includegraphics[width=0.49\textwidth]{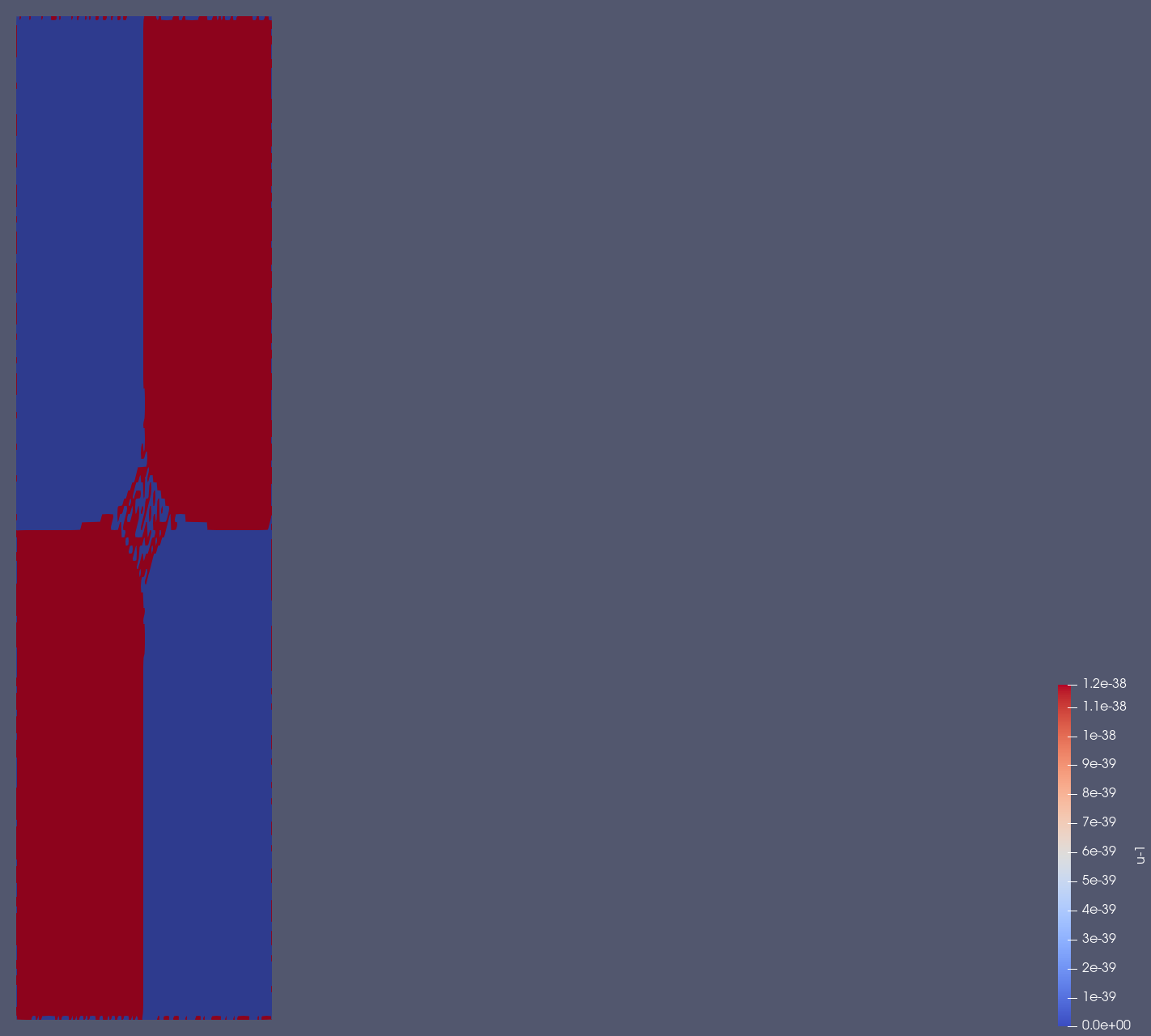}
\caption{Numerical steady state solution of the real
(left) and imaginary (right) parts of the density matrix in convenient coordinates under a quartic potential after an evolution time of $t=50.0$, solved by DG. Remark: Color scale differs between right picture and left one.}
\end{figure}

\section{Conclusions}

Work  has been presented regarding the setup of DG numerical schemes applied to a transformed Master Equation, obtained as the Fourier transform of the WFP model for open quantum systems. The Fourier transformation was applied over the WFP equation in order to reduce the computational cost associated with the pseudo-differential integral operator appearing  in WFP. The model has been expressed as a system of equations by decomposing it into its real and imaginary parts (when expressing the density matrix in terms of the position basis). Given the $\eta$-transport and $x$-related gradient in the diffusion for this problem, the system has been set up so that a DG method can be implemented for the desired numerical solution. 
Numerical simulations have been presented  for the computational study of a benchmark problem such as the case of a harmonic potential, where a comparison between the numerical and analytical steady-state solutions can be performed for a long enough simulation time. 
Further general potentials could be studied in future work for the analysis of perturbations in an uncertainty quantification setting (or to also consider self-consistent interaction effects between the agents under consideration), 
as well as the case of $d>1$ 
for studying the
interaction of a system with a noisy environment, as in the Noisy Intermediate Scale Quantum (NISQ)
devices regime in higher dimensions $d$.

\section*{Acknowledgments}

Start-up funds support from UTSA 
is gratefully acknowledged by the author.

\bibliographystyle{siam}
\bibliography{mybibliography}

\begin{thebibliography}{10}

\bibitem{AlnaesEtal2015}
{\sc M.~S. Alnaes, J.~Blechta, J.~Hake, A.~Johansson, B.~Kehlet, A.~Logg,
  C.~Richardson, J.~Ring, M.~E. Rognes, and G.~N. Wells}, {\em The {FEniCS}
  project version 1.5}, Archive of Numerical Software, 3 (2015).

\bibitem{AlnaesEtal2014}
{\sc M.~S. Alnaes, A.~Logg, K.~B. Ølgaard, M.~E. Rognes, and G.~N. Wells},
  {\em Unified form language: A domain-specific language for weak formulations
  of partial differential equations}, {ACM} Transactions on Mathematical
  Software, 40 (2014).

\bibitem{Arnold}
{\sc A.~ARNOLD, F.~FAGNOLA, and L.~NEUMANN}, {\em QUANTUM FOKKER-PLANCK MODELS:
  THE LINDBLAD AND WIGNER APPROACHES}, vol.~23, World Scientific, 2008,
  ch.~N/A.

\bibitem{doi:10.1137/0732084}
{\sc A.~Arnold and C.~Ringhofer}, {\em Operator splitting methods applied to
  spectral discretizations of quantum transport equations}, SIAM Journal on
  Numerical Analysis, 32 (1995), pp.~1876--1894.

\bibitem{55c98552-dbda-399e-bf25-fc085941dc30}
{\sc A.~Arnold and C.~Ringhofer}, {\em An operator splitting method for the
  wigner-poisson problem}, SIAM Journal on Numerical Analysis, 33 (1996),
  pp.~1622--1643.

\bibitem{HronKarlin}
{\sc J.~Blechta and J.~Hron}, {\em Fenics tutorial 1.5.0 documentation}.
\newblock
  \url{https://www.karlin.mff.cuni.cz/~hron/fenics-tutorial/index.html},
  2014--2015.

\bibitem{demeio2003splitting}
{\sc L.~Demeio}, {\em Splitting-scheme solution of the collisionless wigner
  equation with non-parabolic band profile}, Journal of Computational
  Electronics, 2 (2003), pp.~313--316.

\bibitem{PhysRevResearch.6.033147}
{\sc Z.~Ding, C.-F. Chen, and L.~Lin}, {\em Single-ancilla ground state
  preparation via lindbladians}, Phys. Rev. Res., 6 (2024), p.~033147.

\bibitem{doi:10.1142/S0217984994000248}
{\sc G.~D’ARIANO, C.~MACCHIAVELLO, and S.~MORONI}, {\em On the monte carlo
  simulation approach to fokker-planck equations in quantum optics}, Modern
  Physics Letters B, 08 (1994), pp.~239--246.

\bibitem{Ferraro}
{\sc A.~Ferraro, S.~Olivares, and M.~G.~A. Paris}, {\em Gaussian states in
  continuous variable quantum information}, 2005.

\bibitem{RevModPhys.62.745}
{\sc W.~R. Frensley}, {\em Boundary conditions for open quantum systems driven
  far from equilibrium}, Rev. Mod. Phys., 62 (1990), pp.~745--791.

\bibitem{Gamba}
{\sc I.~Gamba, M.~P. Gualdani, and R.~W. Sharp}, {\em {An adaptable
  discontinuous Galerkin scheme for the Wigner-Fokker-Planck equation}},
  Communications in Mathematical Sciences, 7 (2009), pp.~635 -- 664.

\bibitem{GaniSchulzDGquantumLiouville}
{\sc V.~Ganiu and D.~Schulz}, {\em Application of discontinuous galerkin
  methods onto quantum-liouville type equations", international workshop on
  computational nanotechnology}, International Workshop on Computational
  Nanotechnology (IWCN),  (2023).

\bibitem{GoudonSDWigner}
{\sc T.~Goudon}, {\em Analysis of a semidiscrete version of the wigner
  equation}, SIAM J. Numerical Analysis, 40 (2002), pp.~2007--2025.

\bibitem{Iserles}
{\sc A.~Iserles}, {\em A First Course in the Numerical Analysis of Differential
  Equations}, Cambridge Texts in Applied Mathematics, Cambridge University
  Press, 2~ed., 2008.

\bibitem{Juengel}
{\sc A.~J{\"u}ngel}, {\em Transport Equations for Semiconductors}, Lecture
  Notes in Physics, Springer, 2009.

\bibitem{Kirby2004}
{\sc R.~C. Kirby}, {\em Algorithm 839: {FIAT,} a new paradigm for computing
  finite element basis functions}, {ACM} Transactions on Mathematical Software,
  30 (2004), pp.~{502--516}.

\bibitem{kirby2010}
\leavevmode\vrule height 2pt depth -1.6pt width 23pt, {\em {FIAT:} numerical
  construction of finite element basis functions}, in Automated Solution of
  Differential Equations by the Finite Element Method, A.~Logg, K.~Mardal, and
  G.~N. Wells, eds., vol.~84 of Lecture Notes in Computational Science and
  Engineering, Springer, 2012, ch.~13.

\bibitem{KirbyLogg2006}
{\sc R.~C. Kirby and A.~Logg}, {\em A compiler for variational forms}, {ACM}
  Transactions on Mathematical Software, 32 (2006).

\bibitem{liu2016entropy}
{\sc H.~Liu and Z.~Wang}, {\em An entropy satisfying discontinuous galerkin
  method for nonlinear fokker--planck equations}, Journal of Scientific
  Computing, 68 (2016), pp.~1217--1240.

\bibitem{LoggEtal2012}
{\sc A.~Logg, K.~Mardal, G.~N. Wells, et~al.}, {\em Automated Solution of
  Differential Equations by the Finite Element Method}, Springer, 2012.

\bibitem{LoggWells2010}
{\sc A.~Logg and G.~N. Wells}, {\em {DOLFIN:} automated finite element
  computing}, {ACM} Transactions on Mathematical Software, 37 (2010).

\bibitem{LoggEtal_10_2012}
{\sc A.~Logg, G.~N. Wells, and J.~Hake}, {\em {DOLFIN:} a {C++/Python} finite
  element library}, in Automated Solution of Differential Equations by the
  Finite Element Method, A.~Logg, K.~Mardal, and G.~N. Wells, eds., vol.~84 of
  Lecture Notes in Computational Science and Engineering, Springer, 2012,
  ch.~10.

\bibitem{LoggEtal_11_2012}
{\sc A.~Logg, K.~B. Ølgaard, M.~E. Rognes, and G.~N. Wells}, {\em {FFC:} the
  {FEniCS} form compiler}, in Automated Solution of Differential Equations by
  the Finite Element Method, A.~Logg, K.~Mardal, and G.~N. Wells, eds., vol.~84
  of Lecture Notes in Computational Science and Engineering, Springer, 2012,
  ch.~11.

\bibitem{lohrengel:hal-00210604}
{\sc S.~Lohrengel and T.~Goudon}, {\em {Discrete model for quantum transport in
  semi-conductor devices}}, {Transport Theory and Statistical Physics}, 31
  (2002), pp.~471--490.

\bibitem{MikeIke}
{\sc M.~A. Nielsen and I.~L. Chuang}, {\em Quantum Computation and Quantum
  Information: 10th Anniversary Edition}, Cambridge University Press, 2010.

\bibitem{IJNAM6533}
{\sc J.~Proft and B.~Riviere}, {\em Discontinuous galerkin methods for
  convection-diffusion equations for varying and vanishing diffusivity},
  International Journal of Numerical Analysis and Modeling, 6 (2009).

\bibitem{12c710cb-09d6-3648-9a11-c65d0b7e8f85}
{\sc C.~Ringhofer}, {\em A spectral method for the numerical simulation of
  quantum tunneling phenomena}, SIAM Journal on Numerical Analysis, 27 (1990),
  pp.~32--50.

\bibitem{06e29f02-3d45-3e6b-93aa-17b8a7d51143}
\leavevmode\vrule height 2pt depth -1.6pt width 23pt, {\em A spectral
  collocation technique for the solution of the wigner-poisson problem}, SIAM
  Journal on Numerical Analysis, 29 (1992), pp.~679--700.

\bibitem{RinghoferConvergence}
{\sc C.~Ringhofer}, {\em On the convergence of spectral methods for the
  wigner-poisson problem}, Mathematical Models and Methods in Applied Sciences,
  02 (2011).

\bibitem{Riviere}
{\sc B.~Rivière, M.~F. Wheeler, and V.~Girault}, {\em A priori error estimates
  for finite element methods based on discontinuous approximation spaces for
  elliptic problems}, SIAM Journal on Numerical Analysis, 39 (2002),
  pp.~902--931.

\bibitem{BasixJoss}
{\sc M.~W. Scroggs, I.~A. Baratta, C.~N. Richardson, and G.~N. Wells}, {\em
  Basix: a runtime finite element basis evaluation library}, Journal of Open
  Source Software, 7 (2022), p.~3982.

\bibitem{ScroggsEtal2022}
{\sc M.~W. Scroggs, J.~S. Dokken, C.~N. Richardson, and G.~N. Wells}, {\em
  Construction of arbitrary order finite element degree-of-freedom maps on
  polygonal and polyhedral cell meshes}, ACM Transactions on Mathematical
  Software,  (2022).
\newblock To appear.

\bibitem{Shao_Lu_Cai_2011}
{\sc S.~Shao, T.~Lu, and W.~Cai}, {\em Adaptive conservative cell average
  spectral element methods for transient wigner equation in quantum transport},
  Communications in Computational Physics, 9 (2011), p.~711–739.

\bibitem{1202616}
{\sc L.~Shifren, C.~Ringhofer, and D.~Ferry}, {\em A wigner function-based
  quantum ensemble monte carlo study of a resonant tunneling diode}, IEEE
  Transactions on Electron Devices, 50 (2003), pp.~769--773.

\bibitem{605aac607d794ea39022c2cd16fd7cc8}
{\sc C.~Sparber, J.~Carrillo, J.~Dolbeault, and P.~Markowich}, {\em On the
  long-time behavior of the quantum fokker-planck equation}, Monatshefte fur
  Mathematik, 141 (2004), pp.~237--257.

\bibitem{Suh1991NumericalSO}
{\sc N.~D. Suh, M.~R. Feix, and P.~Bertrand}, {\em Numerical simulation of the
  quantum liouville-poisson system}, Journal of Computational Physics, 94
  (1991), pp.~403--418.

\bibitem{OlgaardWells2010}
{\sc K.~B. Ølgaard and G.~N. Wells}, {\em Optimisations for quadrature
  representations of finite element tensors through automated code generation},
  {ACM} Transactions on Mathematical Software, 37 (2010).

\end{thebibliography}
%

\end{document}